\documentclass[12pt,preprint]{aastex} 
 
\usepackage{epsfig} 

\slugcomment{Revised version after referee report, 2/4/2003} 
 
\shorttitle{The intracluster stellar population in N-body simulations} 
\shortauthors{Napolitano et al.} 
 
\begin{document}

\title{Intracluster stellar population properties from N-body cosmological simulations -- I. Constraints at $z=0$} 
 
\author{Nicola R. Napolitano\altaffilmark{1,2} \& Maurilio Pannella\altaffilmark{1,3}}

\author{Magda Arnaboldi\altaffilmark{4,1}, Ortwin
Gerhard\altaffilmark{5}, J. Alfonso L.  Aguerri\altaffilmark{6},
Kenneth C. Freeman\altaffilmark{7}, Massimo
Capaccioli\altaffilmark{1,8}}
 
\and 
 
\author{Sebastiano Ghigna\altaffilmark{9}, Fabio
Governato\altaffilmark{10,11}, Tom Quinn\altaffilmark{10}, Joachim
Stadel\altaffilmark{12}}
 
\altaffiltext{1}{INAF--Astronomical Observatory of Capodimonte, Via Moiariello 16, I-80131 Naples, Italy} 
\altaffiltext{2}{{\em Present address}: Kapteyn Astronomical Institute, Landleven, 12, PostBus 9700, Groningen, Netherlands} 
\altaffiltext{3}{{\em Present address}: MPE--Max Planck Institut fuer Extraterrestrische Physik, Giessenbachstrasse D-85748 Garching b. Muenchen, Germany} 
\altaffiltext{4}{INAF--Astronomical Observatory of Pino Torinese, Via Osservatorio 20, I-10025 Pino Torinese, Italy} 
\altaffiltext{5}{Astronomisches Institut der Universit\"at Basel, Venusstrasse 7, Binningen, Switzerland} 
\altaffiltext{6}{Instituto de Astrof\'{\i}sica de Canarias. C/ V\'{\i}a L\'actea s/n. 38200 La Laguna. Spain} 
\altaffiltext{7}{RSAA, Mt. Stromlo Observatory, Weston Creek P.O., ACT 2611} 
\altaffiltext{8}{Dept. of Physical Sciences, University ``Federico II'', Naples, Italy}  
\altaffiltext{9}{Universita' di Milano Bicocca, Milan, Italy} 
\altaffiltext{10}{Astronomy Dept, University of Washington, Box 351580, Seattle, WA} 
\altaffiltext{11}{INAF--Osservatorio  Astronomico di Brera, Milan, Italy} 
\altaffiltext{12}{Institute for Theoretical Physics, University of Zurich, Winterthurerstrasse 190, Switzerland} 
\clearpage 
 
\begin{abstract} 
 
We use a high resolution collisionless simulation of a Virgo--like 
cluster in a $\Lambda$CDM cosmology to determine the velocity and 
clustering properties of the diffuse stellar component in the 
intracluster region at the present epoch.  The simulated cluster 
builds up hierarchically and tidal interactions between member 
galaxies and the cluster potential produce a diffuse stellar component 
free-flying in the intracluster medium.  Here we adopt an empirical 
scheme to identify tracers of the stellar component in the simulation 
and hence study its properties.  We find that at $z=0$ the 
intracluster stellar light is mostly unrelaxed in velocity space and 
clustered in structures whose typical clustering radii are about 50 
kpc at R=400--500 kpc from the cluster center, and predict the radial velocity 
distribution expected in spectroscopic follow-up surveys.  
Finally, we compare the spatial clustering in the simulation with the properties of the Virgo intracluster stellar population, as traced by ongoing intracluster 
planetary nebulae surveys in Virgo. The preliminary results indicate a substantial agreement with the observed clustering properties of the diffuse stellar population in Virgo.  
 
\end{abstract}


\section{Introduction} 
The diffuse light in nearby galaxy clusters is now clearly detected \citep{fel02} and its properties can be mapped via the Intracluster Planetary 
Nebulae (ICPNe hereafter) out to large radii 
\citep{aa96,theu97,cia98,fel98,papI,ok02,papIII}. Recently, a complete 
sample of ICPNe identified by the [OIII] $\lambda$ 5007\AA\ emission 
line in different fields in Virgo has been used in \citet[][in 
preparation]{papII} to study the 2D spatial distribution of the 
diffuse intracluster stellar population. 
 
While the presence of an intracluster stellar population (ICSP) has 
now been clearly demonstrated, its origin, time of formation, spatial 
distribution, and even its mass compared to that of the total 
stellar population are still largely unknown.  Different cluster 
formation mechanisms might predict, at the present epoch, different 
spatial distributions or distribution functions for the ICSP that can 
be tested against the 2-dimensional (2D) projected distribution of 
ICPNe and their radial velocities. 
 
\citet{me84} proposed a model where the morphology distribution of 
galaxies in clusters is fixed during the cluster collapse and changes 
very little afterward. In this picture, tidal interactions have very 
limited effects on galaxies, with only those galaxies that are located 
deep in the cluster potential being affected.  In such a model, the 
ICSP is removed from galaxies early during the cluster collapse, and 
its distribution is predicted to follow closely that of galaxies. 
 
In the currently favored hierarchical clustering scenario, fast
encounters and tidal interactions within the cluster potential are the
main players of the morphological evolution of galaxies in clusters.
``Galaxy harassment'' \citep[][]{moo96} and ``tidal stirring''
\citep{may01} cause a significant fraction of the stellar component in
individual galaxies to be stripped and dispersed within the cluster in
a few dynamical times. If the time scale for significant phase--mixing
is longer than a Hubble time (or of the order of few cluster internal
dynamical times), then the ICSP fraction should still be located in
long streams along the orbits of the parent galaxies \citep[as perhaps
observed in Coma and Hydra,][]{calc00}. Detections of substructure in
phase space \citep[see analog in the Milky Way,][]{hel01} would be a
clear sign of harassment as the origin of the ICSP. Dynamical
properties of the ICSP are an important key in the puzzle of cluster
evolution.
 
Until now, N-Body simulations have been used to investigate various
observables in clusters of galaxies such as the density profiles
\citep{nav97, moo98a}, projected galaxy number density \citep{moo99c},
dynamical properties of galaxies \citep{gov01a} and X-ray properties
of the intracluster medium \citep{Borg01}. \citet{dub98} studied
the properties of the cluster stellar population in a CDM cosmological
simulation, focussing on the origin of brightest cluster galaxies
(BCGs). Semi-analytical models were used to study the properties of
the baryonic component, e.g. the galaxy luminosity functions,
Tully-Fisher and Faber-Jackson relations, number counts, the
distribution of morphology, color and size, clustering strengths, and
velocity dispersion profiles
\citep{kau94,col94,guid98,mo98,gov98,vdB00,dia01,spri01}. However,
this powerful approach is not able to follow in detail the dynamical
evolution of the stellar component, once stripped from the parent
galaxies.

We use here a cosmological N-body simulation to 
study the properties of the intracluster stellar population in a 
Virgo--like cluster at $z=0$, in a $\Lambda$CDM cosmology. N-body 
simulations for the dark matter component\ allow a larger number 
of particles and a larger dynamical range than simulations including 
an explicit treatment of hydrodynamics and star formation processes.
Because stars are expected to form as gas cools at the bottom of the 
potential wells of dark matter halos \citep{wr78}, the simulated 
particles in the very high density regions (i.e. the center of DM 
halos) are likely to be good tracers of the stellar component.  We 
therefore select as tracers of the stellar population those particles 
within local densities higher than $\sim 10^4$ times the critical 
density at any given time before $z\ge0.25$, which is considered the 
epoch when the star formation stopped in the cluster. 
 
This approach has some interesting advantages, when compared with the 
lower resolution hydrodynamical simulations with an explicit 
treatment of star formation.  Namely we are able to follow the 
dynamical evolution of much smaller galaxies, and the higher numerical 
resolution helps to decrease significantly the numerical noise and the 
two--body relaxation, which can rapidly erase structures in the phase 
space. 
 
The aim of this paper is to study the total amount of intracluster 
stellar light, the clustering properties, and the velocity 
distribution of the intracluster population of stars in a Virgo--like 
cluster in $\Lambda$CDM cosmology. We can then make predictions for 
observables such as the two-point angular correlation function (2PCF, 
hereafter) and the spatial correlation function (both of these 
computed via the LS estimator, see Appendix A) and the line of sight 
(LOS, hereafter) velocity distribution as expected in a present-day 
cluster.  The former will be compared with the data samples now 
becoming available from wide field imaging surveys; the latter will 
soon be constrained by spectroscopical surveys with wide-field 
spectrographs. Linking the observable clustering and velocity 
properties of the ICSP with results from N-body simulations can in 
principle also be used to put constraints on the cluster formation 
epoch. \\ 
 
The paper is organised as follows: in Section 2 we describe the N-body 
cosmological simulations and the schema adopted to identify the ICSP; 
in Section 3 we describe the comparison with the observed datasets and 
the analysis of the velocity and clustering properties of the ICSP 
population; in Section 4 the dynamical characterisation of the 
intracluster stellar population is discussed.  Conclusions are given 
in Section 5.

\section{N-body Cosmological simulation \label{selpro}} 
 
We have analysed a high resolution simulation of the formation of a 
galaxy cluster in $\Lambda$CDM cosmology \citep{gov01b}, with $h_0$=0.7, 
$\sigma_8$=1, $\Omega_0$=0.3, $\Lambda$=0.7. The simulation was done 
using the so-called ``renormalization technique'' \citep{ka93}, with a 
total of 1.5 million particles, in a volume of 100 Mpc on a side. The 
cluster builds up hierarchically and has a typical merging history 
within the adopted cosmology, with a few groups being accreted at 
relatively low $z$.  Tidal interactions between member galaxies and 
the cluster potential produce a diffuse stellar component free flying 
in the intracluster medium.  At the end of the simulation, the cluster 
total mass is $\sim 3 \times 10^{14} M_\odot$, similar to the Virgo 
cluster total mass, and half a million particles are within the 
cluster virial radius; each particle has a mass of $0.506 \times 
10^{9} M_{\odot}$. The spatial resolution is 2.5 kpc, which allows us 
to resolve substructures down to sub L$_{*}$ scales. Several thousands 
of time steps were used to follow the simulation up to the present 
time.  More details of the simulation are given in \citet{gov01b} 
and \citet{Borg01}.

\subsection{Tracing the stellar component\label{ovdens}} 
 
Following \citet{wr78}, the very high density regions at the center of 
DM halos are likely to be good tracers of the stellar component in 
galaxies.  A simple criterion to select such tracers in an N-body 
simulation is via a density threshold; this is easily implemented and 
the implications can be studied in detail.  We therefore select as 
tracers of the stellar population those particles in the simulation 
within local overdensities higher than $\sim 10^4$ times the critical 
density at any given time before $z\ge0.25$, which is considered the 
epoch when the star formation stopped in the cluster. 
 
For a fixed density threshold, the size of such an overdensity 
region varies with redshift, and for a given cosmological model depends mostly on the virial mass and 
concentration \citep[][]{nav97,bull01}, and possibly on the accretion 
history of the dark halos \citep[][]{wech02,zhao02} because of the 
structure growth.  The evolution with redshift of the overdensity 
regions in the growing halos is also determined by the cosmological model 
\citep[see][for a discussion]{nav97,bull01}.  We estimated that the 
typical size of an overdensity region with 12000 times the critical 
density has a linear dimension of about 15 kpc at $z=0$ and 12 kpc at 
$z=3$ for a $M_{vir}=5\times 10^{12} h^{-1} M_{\odot}$ (a typical 
virial mass of a brightest cluster galaxy -- BCG hereafter) and it is 7 kpc at $z=0$ 
and 5 kpc at $z=3$ for a virial mass of $M_{vir}=1.25\times 10^{12} h^{-1} M_{\odot}$ \citep[][]{bull01,wech02,zhao02}. Those particles within 
these scales experience local overdensities which are larger than the 
adopted threshold. 
 
How do these overdensity regions compare with the optical radius of 
galaxies at different redshifts?  Studies on the evolution of the 
linear scales of the luminous parts of galaxies suggest that they are 
a decreasing function of redshift; \citet{nel02} found that the 
effective radius, $r_e$, for their galaxy sample decreases by a factor 
two at $z\sim0.8$ with respect to the local values.  \citet{sim99} 
showed that the total galaxy light is more concentrated at $z\sim1$ 
with respect to local values. 
 
This evidence implies that our simple density threshold 
criterion will select particles in the central regions of dark halos 
whose size is comparable to the galaxy luminous part. Some of these 
particles may be DM particles when selected at high--$z$, i.e $z\ge 
1$, while when selected at lower redshifts, once the overdensity 
regions have reached the same scale of the luminous parts of galaxies, 
they will be most likely tracers of the stellar component and have 
typical stellar M/L ratios. 
 
The particles in the N-body simulation which trace the stellar 
population are thus  selected according to the following procedure:  
we measure the local density around each particle at different 
redshifts ($z=$ 3, 2, 1, 0.5, 0.25).  Then we flag all those particles  
as tracers of the stellar mass which:\\  
i) are found in a local 
overdensity of at least 12000 times the critical density for at least 
one output redshift;\\  
ii) at $z\le$0.5, we remove all those particles  
which are in a 12000 $\rho_{\rm crit}$ overdensity or in the central 
part of the cluster, 
{\it and} that were not marked at any other previous $z$.   
The aim here is to disregard those particles which are situated in the 
cluster center at low $z$, but did not 
belong to any halos at higher $z$, and therefore it is unlikely 
that they trace any stellar component. 
 
\begin{figure} 
\centering 
\epsfig{figure=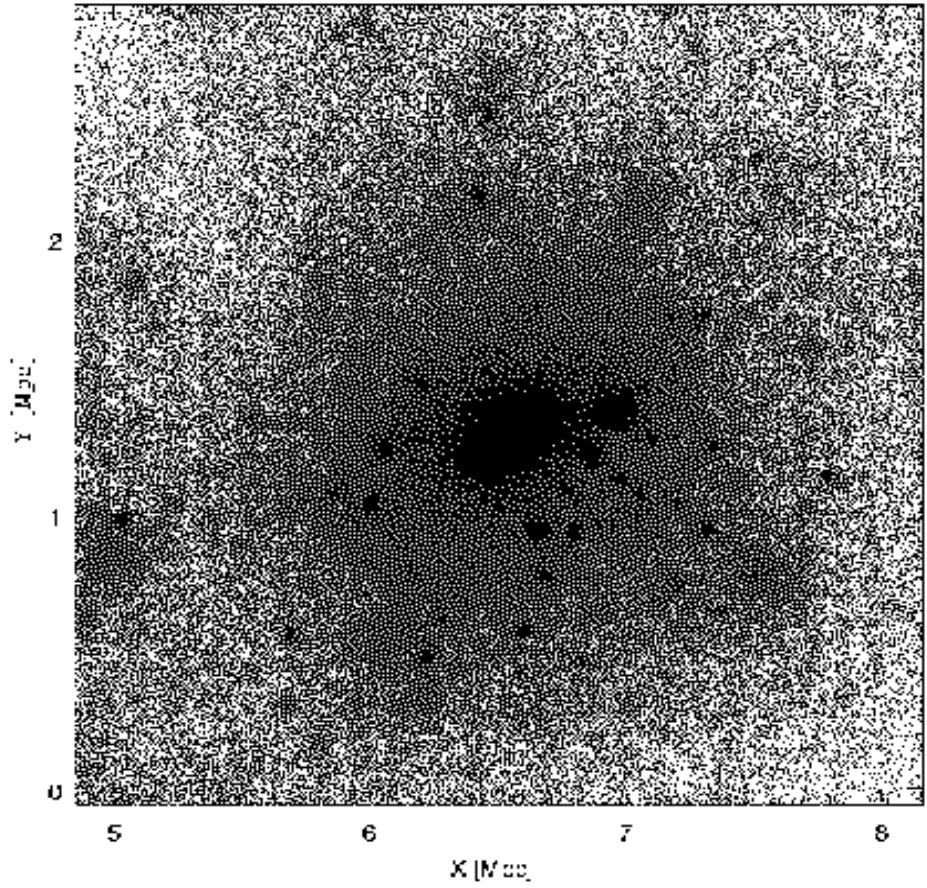,width=18cm,height=22cm} 
\caption{The simulated cluster used in our analysis. In gray we show all the mass particles in the simulation, in black the particles selected according to the criteria discussed in Section \ref{selpro}.} 
\label{rcfi2} 
\end{figure}

All the particles flagged by this procedure are shown in
Figure~\ref{rcfi2}, and they are overplotted to all the
mass--particles in the simulation in the Virgo--like cluster region.
These flagged particles provide us with a subsample of particles from
the simulation, that have spent part of their lifes in the
high-density region of at least one halo at $z>0.5$, and as previously
argued, are good tracers of the baryons that formed stars in these
dark halos. Then the division of the stellar component into galaxies
or intracluster regions at $z =0$ can be easily achieved by selecting
appropriate fields in the simulated data.

\begin{figure} 
\centering 
\epsfig{figure=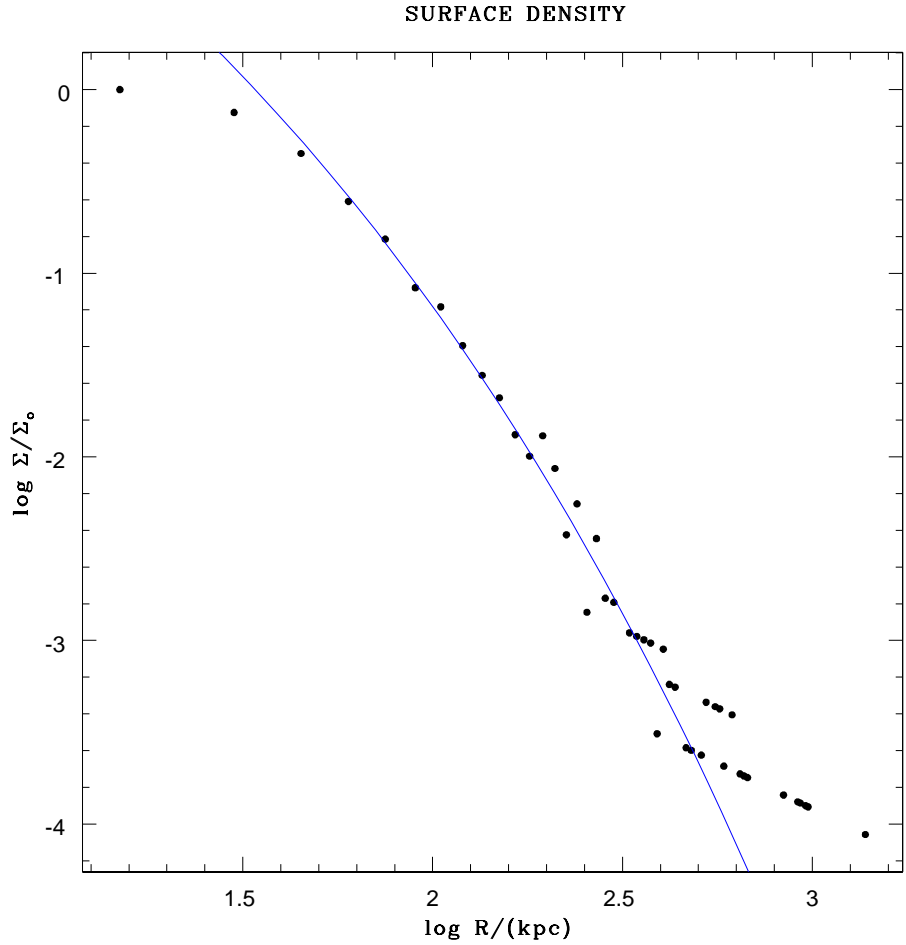,width=13cm,height=13cm} 
\caption{Surface density profile of the selected tracer particles
in a cone free of surrounding subhalos, normalised to the central
value, the full line shows an R$^{1/4}$ profile fitted to the inner
parts. The outer profile indicates that the particle distribution
there is not yet relaxed and some excess over the inner profile is
present.}
\label{cD} 
\end{figure}

The surface density distribution of selected particles is shown
in Figure~\ref{cD}. In producing this figure we have used particles
in a cone which was free of obvious subhalo concentrations, so as to
isolate the smooth component centred on the BCG. In the inner parts,
this surface density profile follow closely an $r^{1/4}$ law; this
result is similar to that of \citet{dub98}, which was based on a
completely different treatment of the luminous component. The outer
profile in Figure~\ref{cD}  shows the unrelaxed nature of the
intracluster component at large radii, and suggests that after
relaxation the density in these parts will have an excess of light in
comparison with the $r^{1/4}$ law.

We have used a simple approach to identify luminous tracer
particles through a density threshold criterion. As Figs.~\ref{rcfi2}
and \ref{cD} show, this leads to a realistic stellar mass
distribution. Because in cosmological dark matter halos with NFW
profiles, relatively few stars in the densest regions come into the
centre on elongated radial orbits from large distances, we expect that
a selection based on binding energy would have produced largely
similar results. We have also neglected the fact that part of the
later intracluster material would be stripped from disks rather than
spheroidal components. Material dispersed into intracluster space from
cold components would be dynamically colder, even though it would have
been typically heated by bar formation prior to stripping.  Even
stripped halos are still cold compared to the cluster velocity
dispersion, however, so they also produce narrow structures in
phase-space. Structures in phase-space originating from cold
components would be even narrower, but not change our main results.

\subsection{The conversion factor to ICPNe} 
Since we want to trace the properties of the ICSP and compare it with
the ICPNe surveys, we need to define a conversion from the
stellar-mass-particles in the N-body simulation to ICPNe, through the
mass-to-light ratio of the stellar population.  We assume that at
$z=0$ the luminosity, mass-to-light ratio, etc., of the
harassed--stellar matter are those of an evolved stellar population
like those of M31.
 
This approach differs with respect to a semi--analytical approach
because there is no Montecarlo realization of the baryonic population
attached to the mass particles \citep[see for
example][]{kau94,col94,dia01,spri01}. All the information on the
stellar population parameters that generate a PN in a galaxy
environment is condensed in the luminosity-specific PN density, which
specifies the number of PNe per unit luminosity.  This quantity
depends on the age and\ metallicity of a stellar population and it is
a measured quantity. Here we adopt the luminosity-specific planetary
nebulae density from M31: $\alpha_{1,B}=9.4\times 10^{-9} PN L_B^{-1}$
\citep[][]{cia89}. This is the best empirically determined value for
an evolved population.  $\alpha_{1,B}$ is the luminosity-specific
planetary nebulae density within 1 mag from the bright cut-off of the
planetary nebulae luminosity function (PNLF). We adopt this value
because the ICPNe samples available in the literature are generally
complete down to about 1 mag from the bright cut-off of their PNLFs.
 
The number of PNe per selected stellar--tracer mass particle (if not 
explicitly specified, we will refer to this selected stellar--tracer 
population as mass particles in what follows) in the cosmological 
simulation is then 
$$n_{PN}/\mathrm{prtcl}= \frac{\alpha_{1,B}\times m_p}{\Gamma_{\mathrm{B}}}= 
\frac{9.4\times10^{-9} \times 0.506\times10^{9}}{\Gamma_{\mathrm{B}}}$$ 
where $m_p=0.506\times10^{9} M_{\odot}$ is the mass of each mass 
particle in the simulation and $\Gamma_{\mathrm{B}}$ is a 
mass-to-light ratio, $\Gamma_{\mathrm{B}}$ = $M/L_B$, for the diffuse 
stellar population from which the ICPNe evolved.  The best estimate 
for $\Gamma_{\mathrm{B}}$ is the value observed in galaxy regions 
where the dynamics is dominated by the luminous component alone.  To 
be consistent with the adopted value of $\alpha_{1,B}$, we take the 
$\Gamma_{\mathrm{B}}$ value for the M31 bulge. \citet{bra91} found 
$\Gamma_{\mathrm{B}} = 6.5\pm0.4$ for the bulge and disk of M31.  With 
these assumptions, we obtain n$_{PN} =0.74$ per mass particle, and 
n$_{PN}=1$ is obtained for a $\Gamma_{\mathrm{B}}$ about 5.  In what 
follows, we will adopt one PN for each mass particle after verifying 
that a $\Gamma_{\mathrm{B}} \sim 5-6 $ is a consistent value for the 
ICSP. 
 
\subsection{Check on the stellar baryon fraction \label{barfra}} 
 
An additional test on our selection of intracluster stars is whether 
the population we have chosen has a reasonable stellar baryon fraction 
in the selected IC fields.  Once a ICSP is selected in an IC field, we 
take its mass to be the total stellar mass in this field, 
$M_{\mathrm{sel}}$. The stellar baryon fractions in this field is 
obtained by dividing $M_{\mathrm{sel}}$ by the total mass 
(dark+stellar), $M_{\mathrm{tot}}$, for all the mass particles in the 
same area. This is the compared with the corresponding number 
estimated for galaxy clusters.

\citet{fuk98} found that the total contribution from stars in 
clusters, i.e. from spheroids, disks and irregular galaxies amounts to 
$\Omega_{\mathrm{stars}}=0.0027$, with an upper limit of about 0.005 
and a minimum of about 0.0015 (for $h_0$=0.7) which, divided by 
$\Omega_{\mathrm{m}}=0.3$ (as in our simulation) gives a stellar 
baryon fraction ranging from 0.005 to 0.016 with a central value of 
$f_{\mathrm{stars}}=\Omega_{\mathrm{stars}}/\Omega_{\mathrm{m}}\sim$0.01. This 
is consistent with the recent estimate of the fraction of cluster 
baryons condensed in stars 
$f_{\mathrm{c,global}}=\Omega_{\mathrm{stars}}/\Omega_{\mathrm{bar}}=0.073 
h$ from \citet{bal01}.  Following these authors in adopting 
$\Omega_{\mathrm{bar}}=0.039 h^{-2}$ \citep[][]{jaf00} and 
$\Omega_{\mathrm{m}}=0.3$, we obtain $f_{\mathrm{stars}}=0.0135$ (for 
$h_0$=0.7).  We assume that the stellar baryon fraction does not 
differ greatly from this mean value at different positions in the 
cluster, and use this estimate in the intracluster fields discussed 
below. 
 
\section{Spatial and velocity structure of the ICSP population} 
\subsection{Field\ selection and comparison strategy with ongoing surveys \label{select}} 
We wish to use the simulated dataset at $z=0$ to predict those 
observables, the 2D distribution and the LOS velocity distribution, 
which can be readily compared with the results from photometric and 
spectroscopic ICPNe surveys. Photometric ICPNe sample are currently 
available \citep[see][]{fel98,papI,papIII,ciar02} and spectroscopic data may 
soon become so. 
 
To facilitate a comparison with the ongoing surveys, we have chosen a 
series of projected regions in our data cube with the typical 
dimensions of the present wide-field-imaging cameras (WFI--like 
fields), 30$' \times$30$'$, at different distances from the simulated 
Virgo--like cluster center. Assuming a distance of 15 Mpc for Virgo 
(as determined from the PNLF of M87), these WFI--like regions cover 
0.131$\times$0.131 Mpc$^2$ in linear scale.  These regions were 
selected at $R=0.2,~0.4,~0.5$ and 0.6 Mpc from the center, which 
corresponds to about $0.5, 1, 1.25 \mbox{ and }1.5$ core radii 
respectively. These fields are placed so as to avoid significant 
sub--halos or overdensities which can be related to cluster 
galaxies. 
 
\begin{figure} 
\epsfig{figure=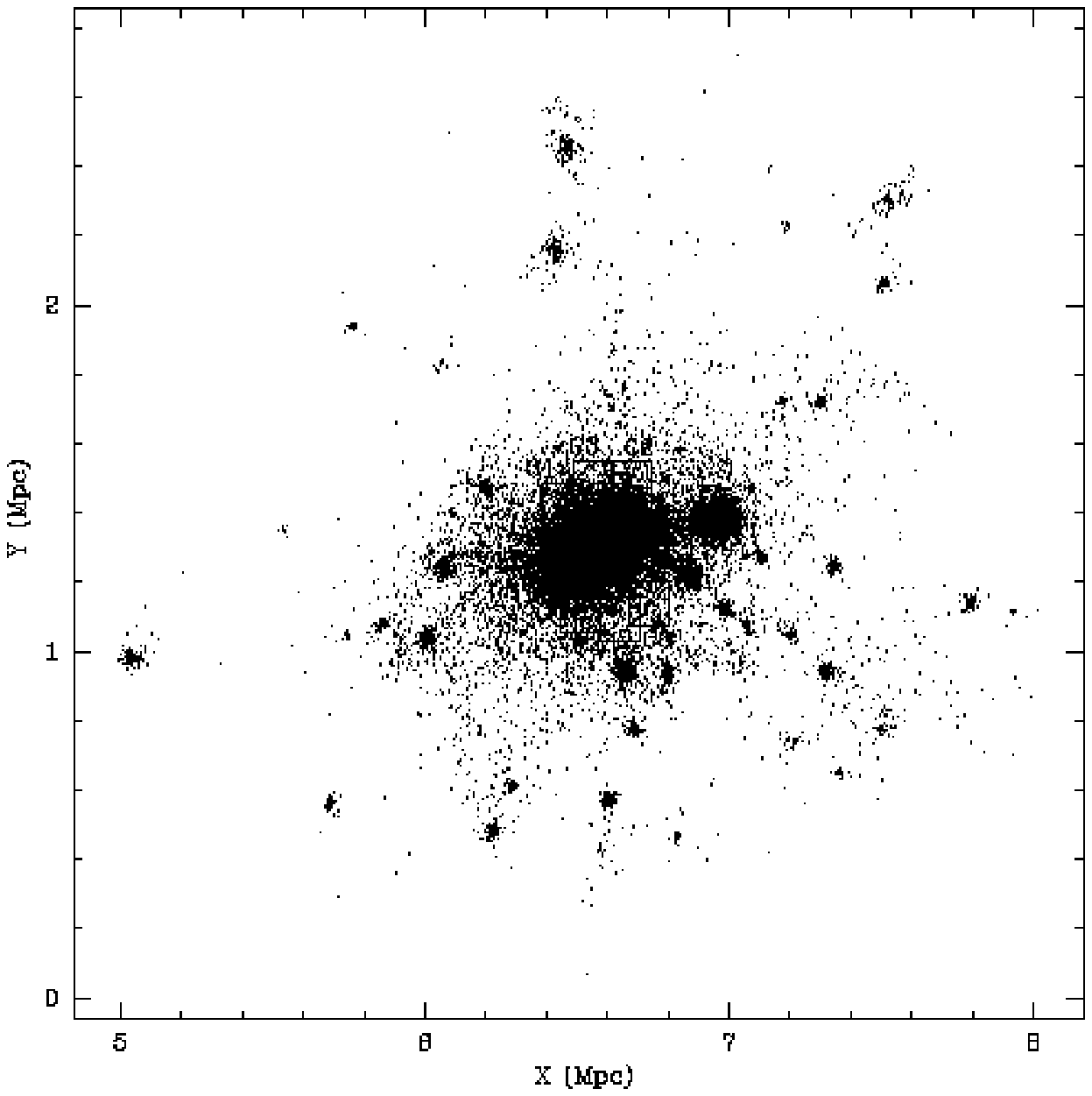,width=8.5cm,height=9.5cm} 
\epsfig{figure=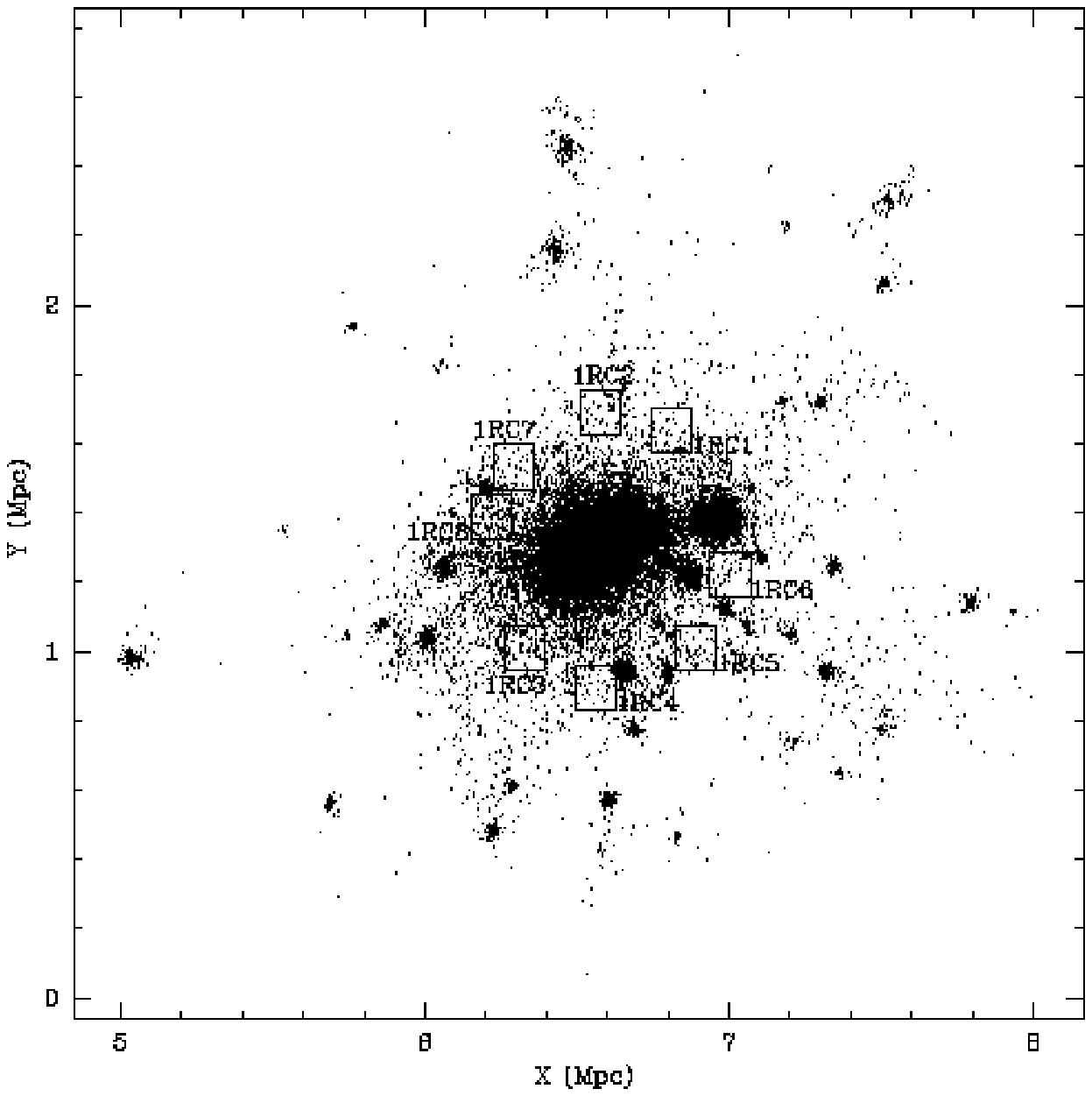,width=8.5cm,height=9.5cm} 
\epsfig{figure=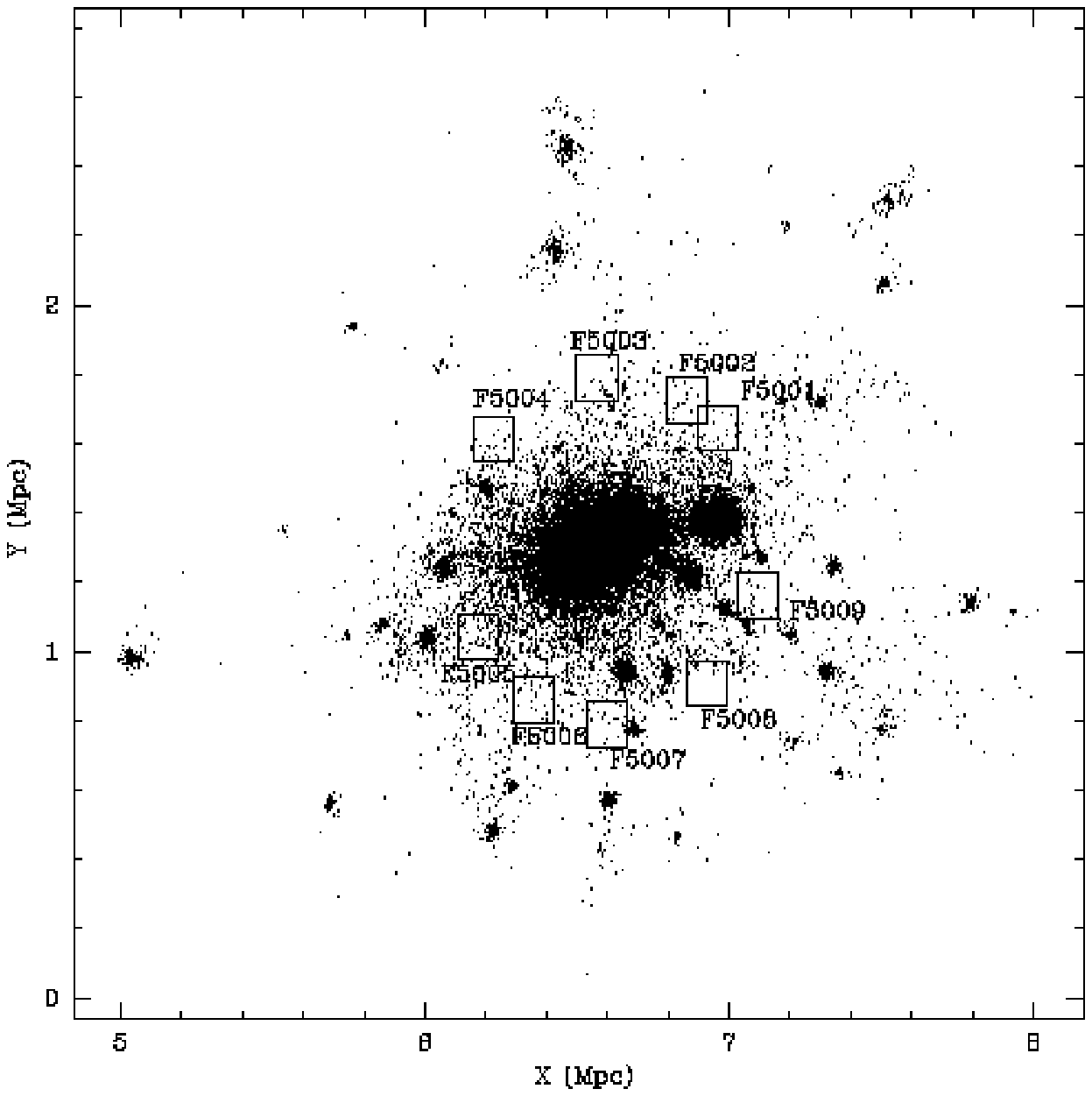,width=8.5cm,height=9.5cm} 
\epsfig{figure=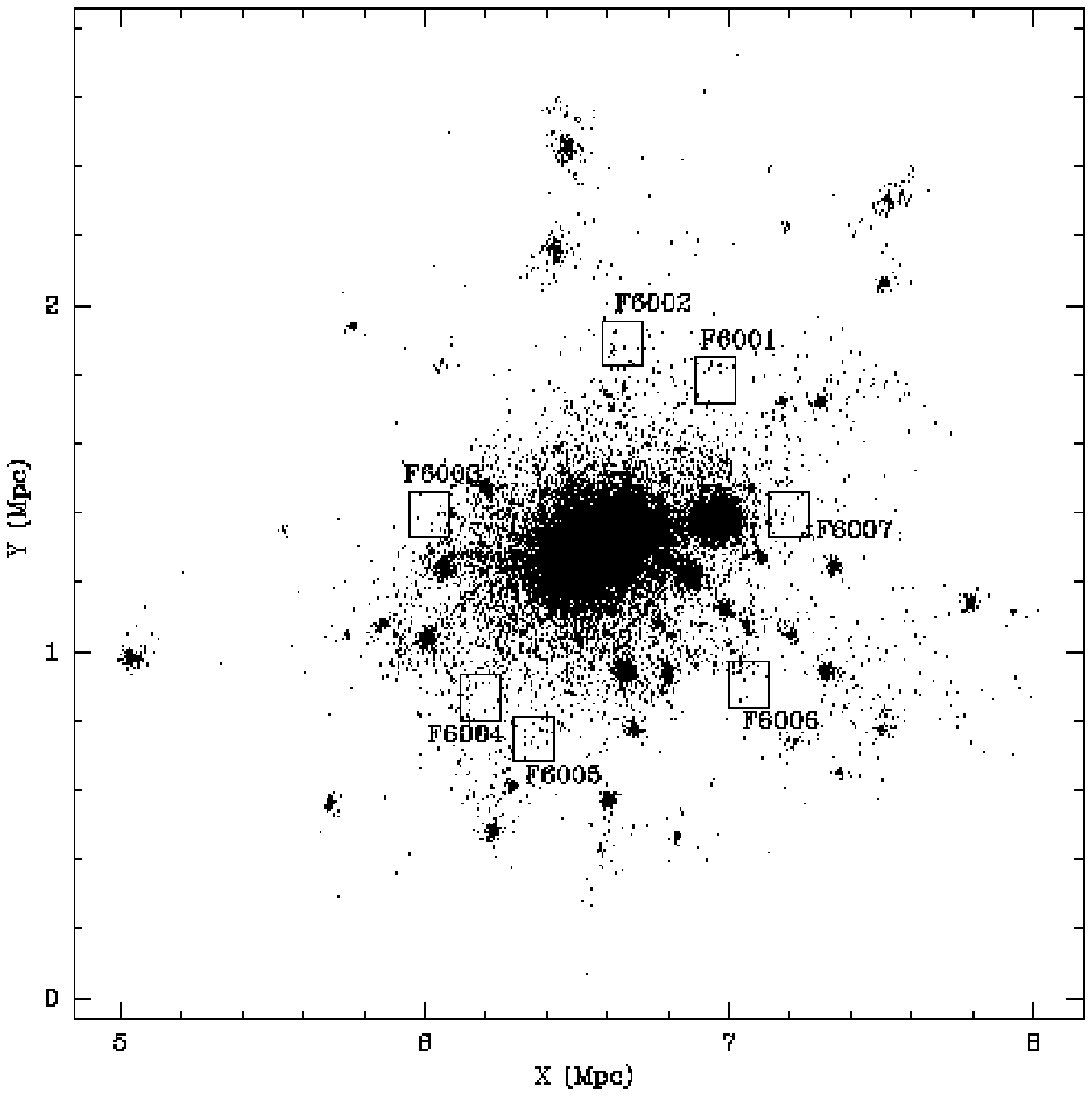,width=8.5cm,height=9.5cm} 
\caption{Here only the stellar tracers in the simulated cluster plus the fields adopted for the analysis are shown: CORE--like fields (top left), RCN1--like fields (top right), $F500$ fields (bottom left) and $F600$ fields (bottom right).} 
\label{rcfi} 
\end{figure}

In what follows, we shall refer to CORE--like and 
RCN1--like for fields placed at $R$=0.2 Mpc and $R$=0.4 Mpc 
respectively (for similarity with the adopted convention in 
\citet{papI} and \citet{papII}).  Fields at $R$=0.5 Mpc and $R$=0.6 
Mpc will be named $F500$ and $F600$ respectively. In Figure 
\ref{rcfi}, we show the fields selected at different distances from 
the cluster center. We have adopted the Z-axis as the line-of-sight 
and the X-Y plane as the plane of the sky\footnote{We have verified 
that the 2D distribution and LOS velocity distribution do not vary on 
average with the viewing angle}, then we have analyzed for each 
field:\\  
1) {\em the M/L ratios} (where information on the observed 
ICSP surface brightness is available from observations) and {\em the 
baryonic stellar fraction} as a check on the ICSP selection;\\  
2) {\em 
the 2D spatial distribution} via the angular two-point correlation function (2PCF) using the \citet{ls93} estimator $\omega(\theta)$, as discussed in the Appendix;\\  
3) {\em the velocity distribution} in order to evaluate the dynamical status of the ICSP:

\subsection{Fields outside of the BCG halo (R$\ge$0.4 Mpc)} 
 
\subsubsection{M/L ratios and baryonic stellar fraction} 
{\em RCN1--like fields.} The global properties of the selected fields 
are summarised in Table~\ref{tabrc}. We determined the M/L ratios for 
the RCN1-like fields in the simulations by computing the local surface 
mass densities from the selected ICSP particles and adopting the surface brightness from \citet{papI}. In most cases, resulting M/L ratios are compatible with 
an ICSP; fields 1RC3 and 1RC8 have larger M/L ratios than for a pure 
stellar population.  Baryonic stellar fractions 
($M_{\mathrm{sel}}/M_{\mathrm{tot}}$ in the fields, see 
Sect. \ref{barfra}) are shown in Table \ref{tabrc}: they are in a 
range of 1-2\% with a mean value of 0.015, in agreement with values 
observed in clusters. 
 
{\em $F500$ and $F600$ fields.} The global properties of the selected 
fields are summarised in Table \ref{tabr5} and \ref{tabr6}.  We do not 
perform a check on the M/L ratios because we have no  surface 
brightness (SB) information; no fields have been 
surveyed so far at such a large distances from the cluster center. The 
mean stellar baryonic fractions are 0.011 and 0.015 respectively, 
consistent with an ICSP population. 
 
\subsubsection{Velocity distributions} 
\begin{figure} 
\vspace{-0.8cm} 
\epsfig{figure=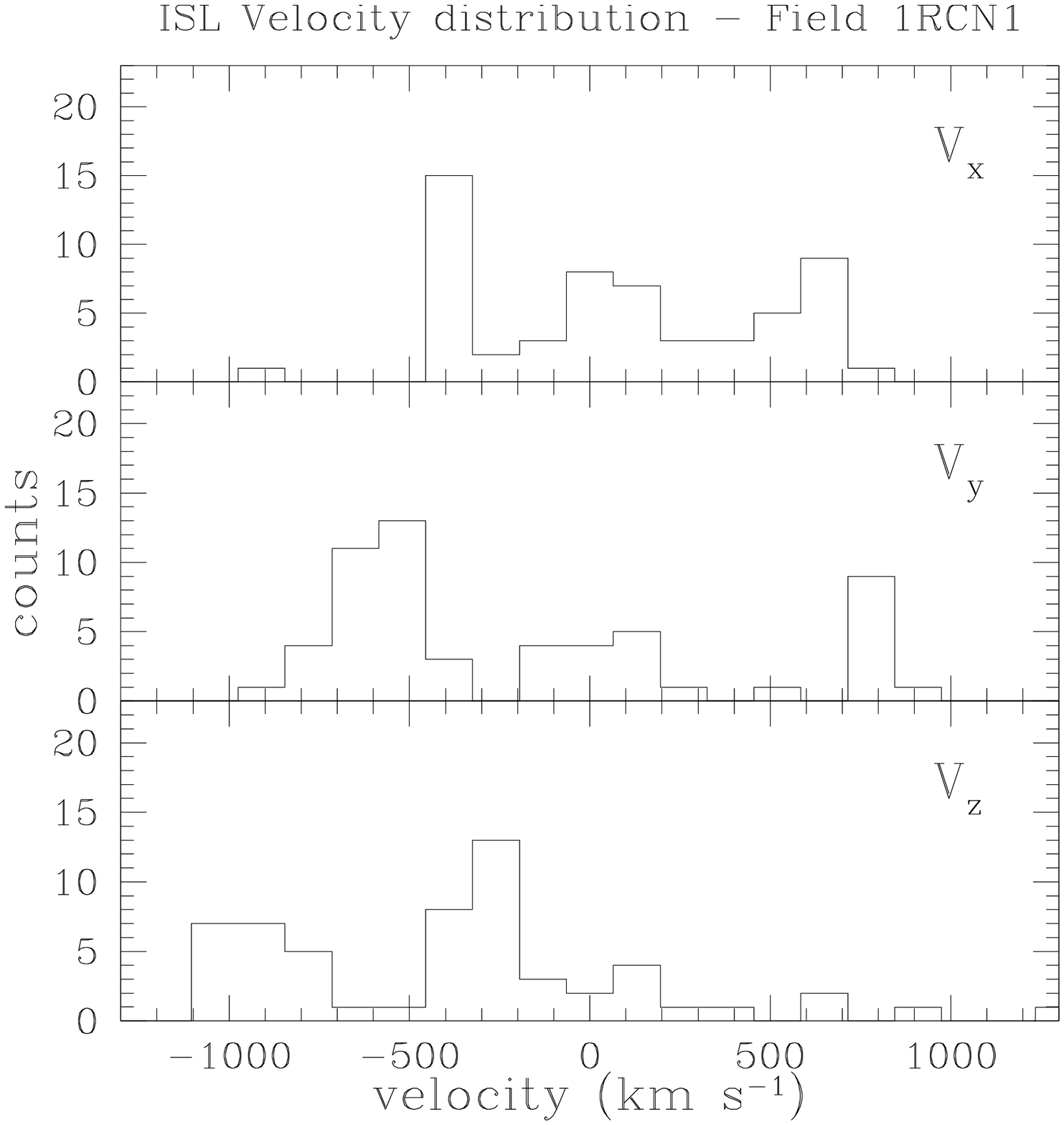,width=4.05cm,height=4.5cm}
\epsfig{figure=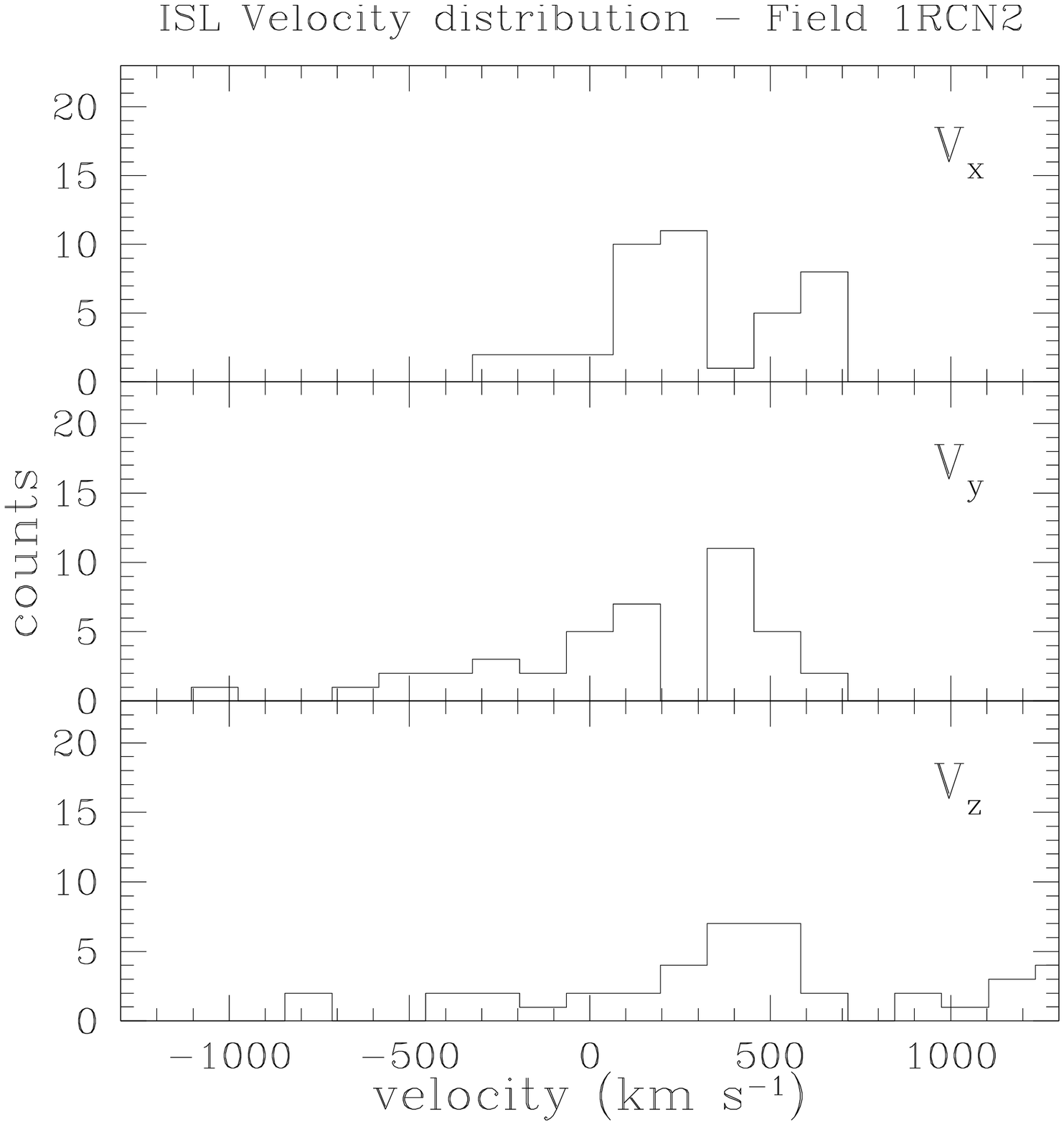,width=4.05cm,height=4.5cm}
\epsfig{figure=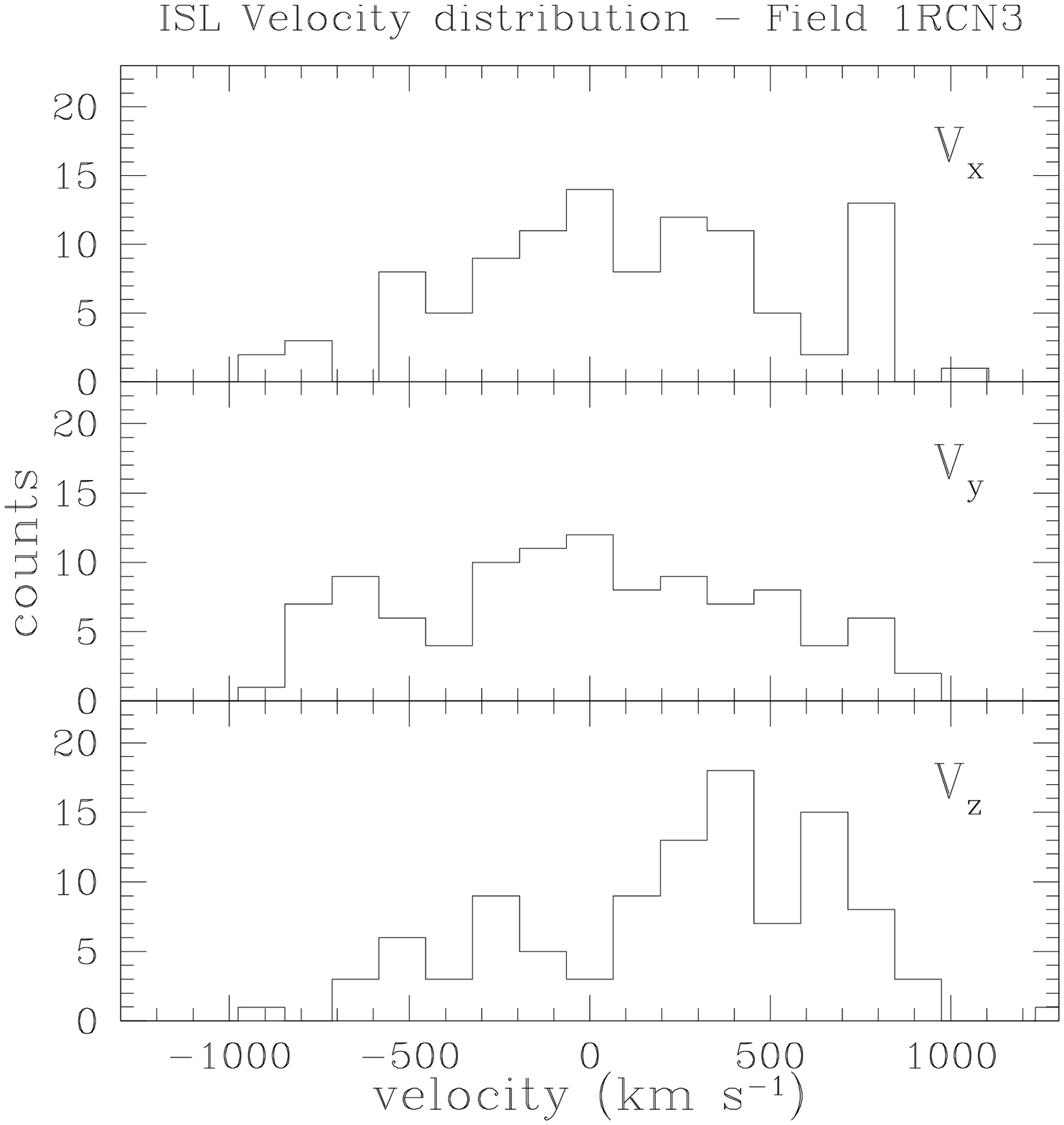,width=4.05cm,height=4.5cm}
\epsfig{figure=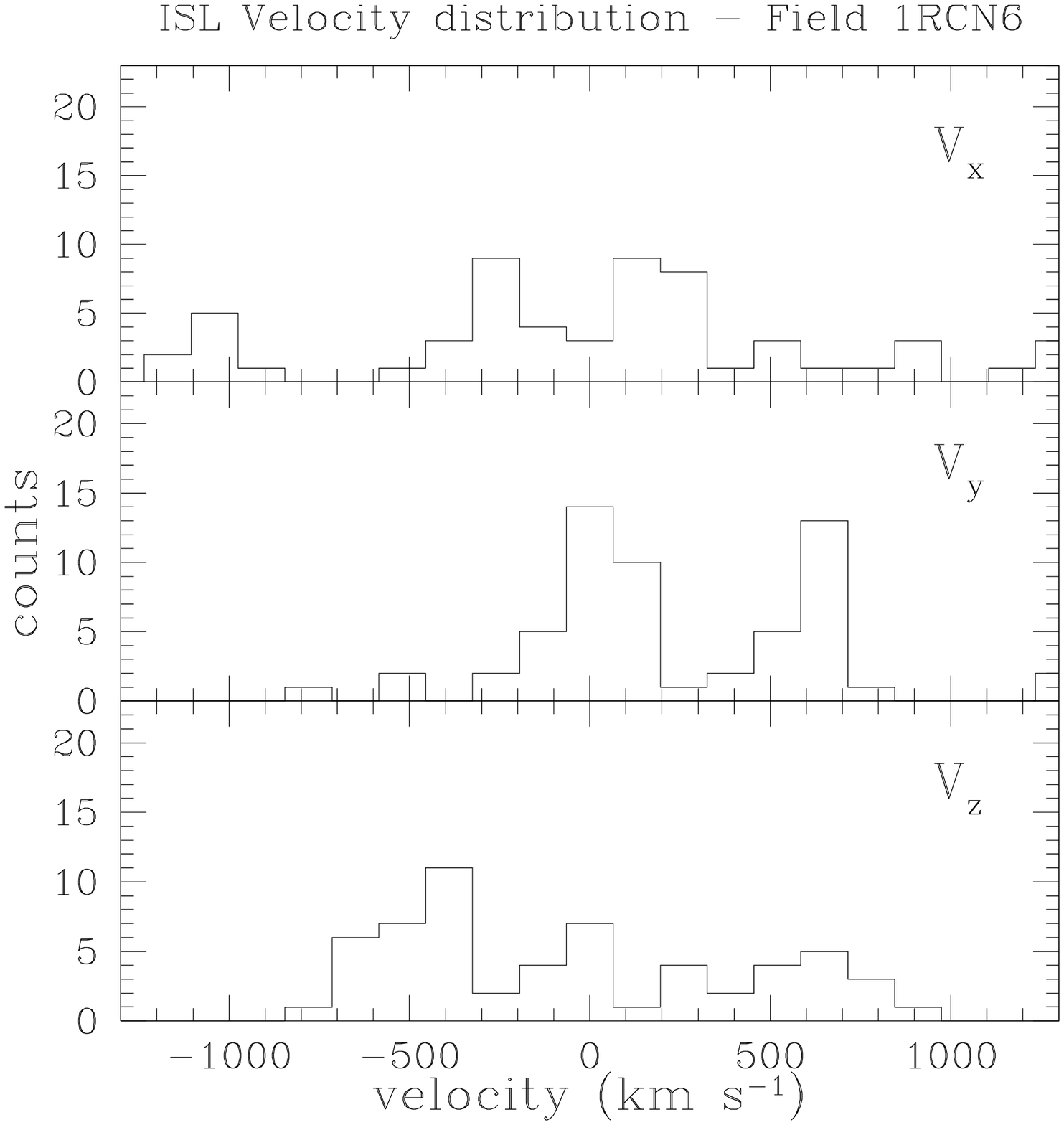,width=4.05cm,height=4.5cm}
\epsfig{figure=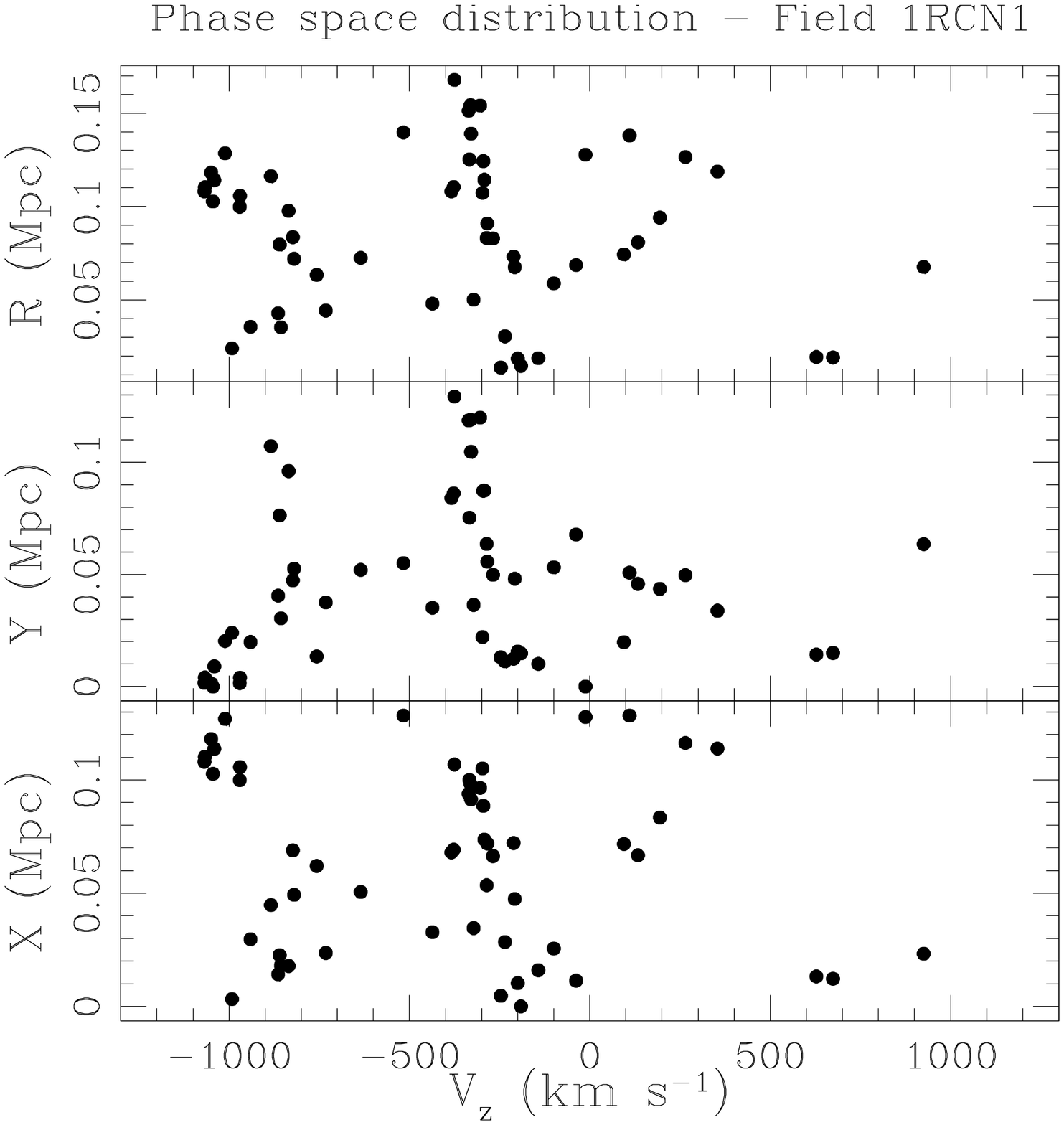,width=4.05cm,height=4.5cm}
\epsfig{figure=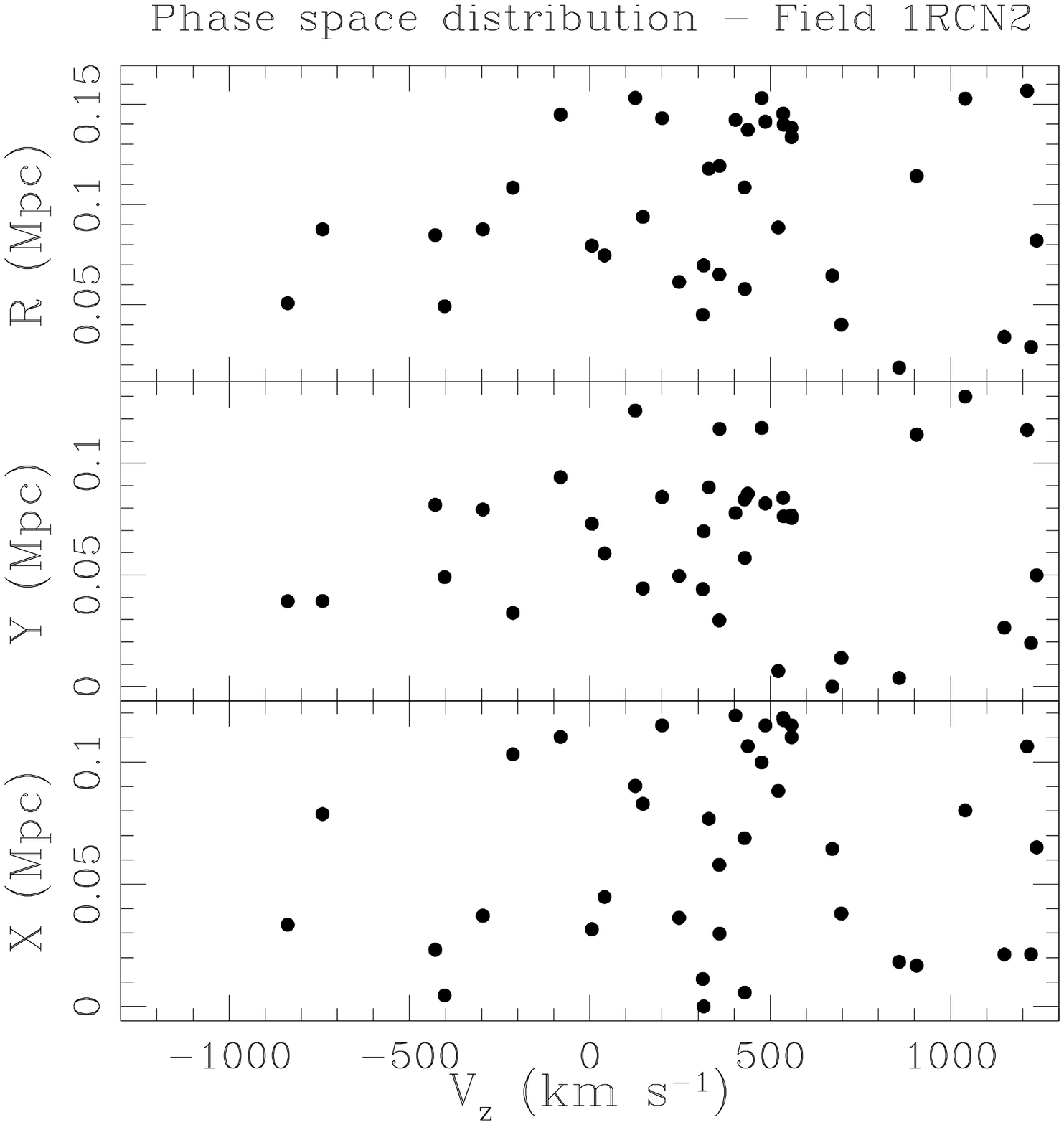,width=4.05cm,height=4.5cm}
\epsfig{figure=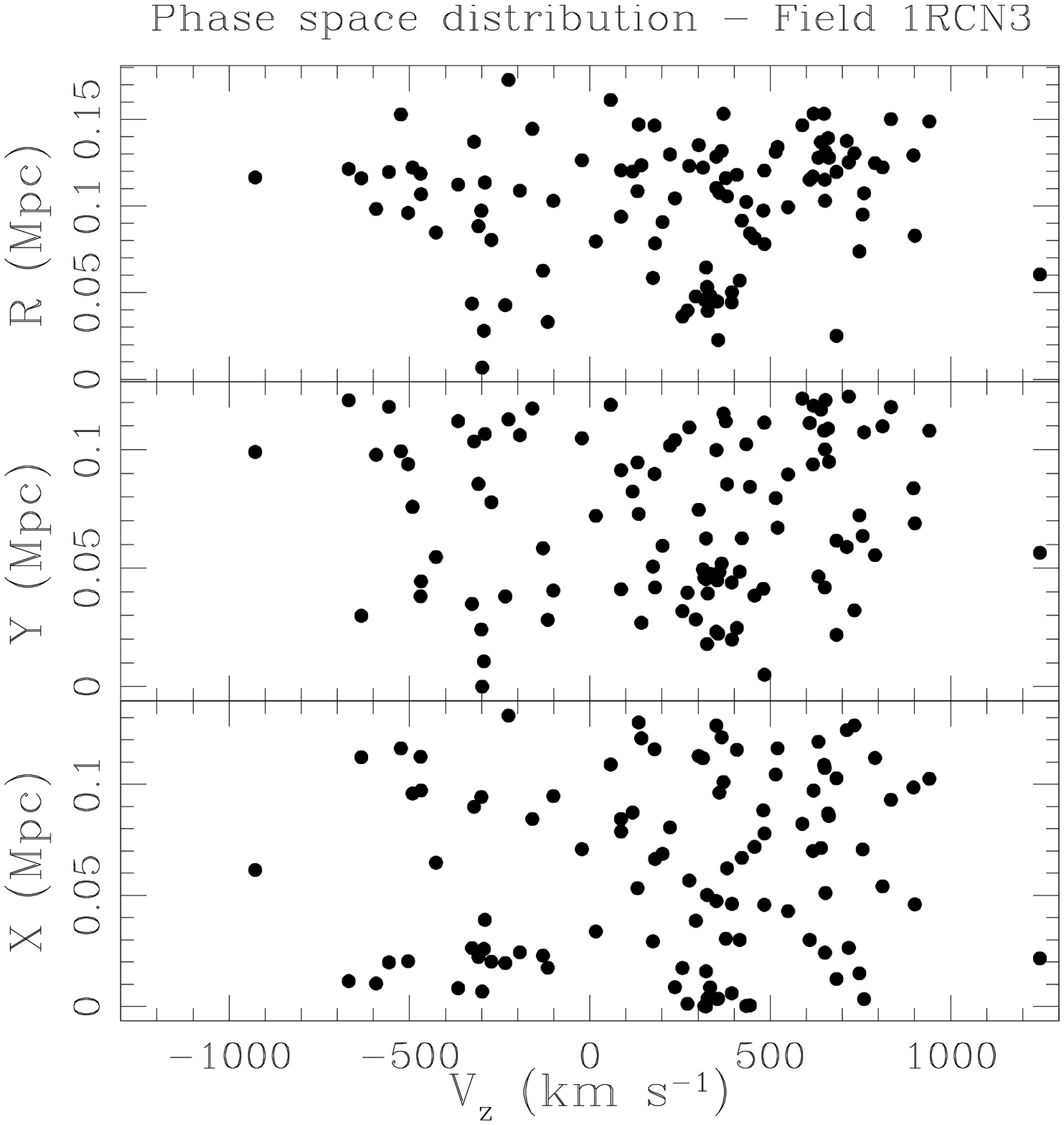,width=4.05cm,height=4.5cm}
\epsfig{figure=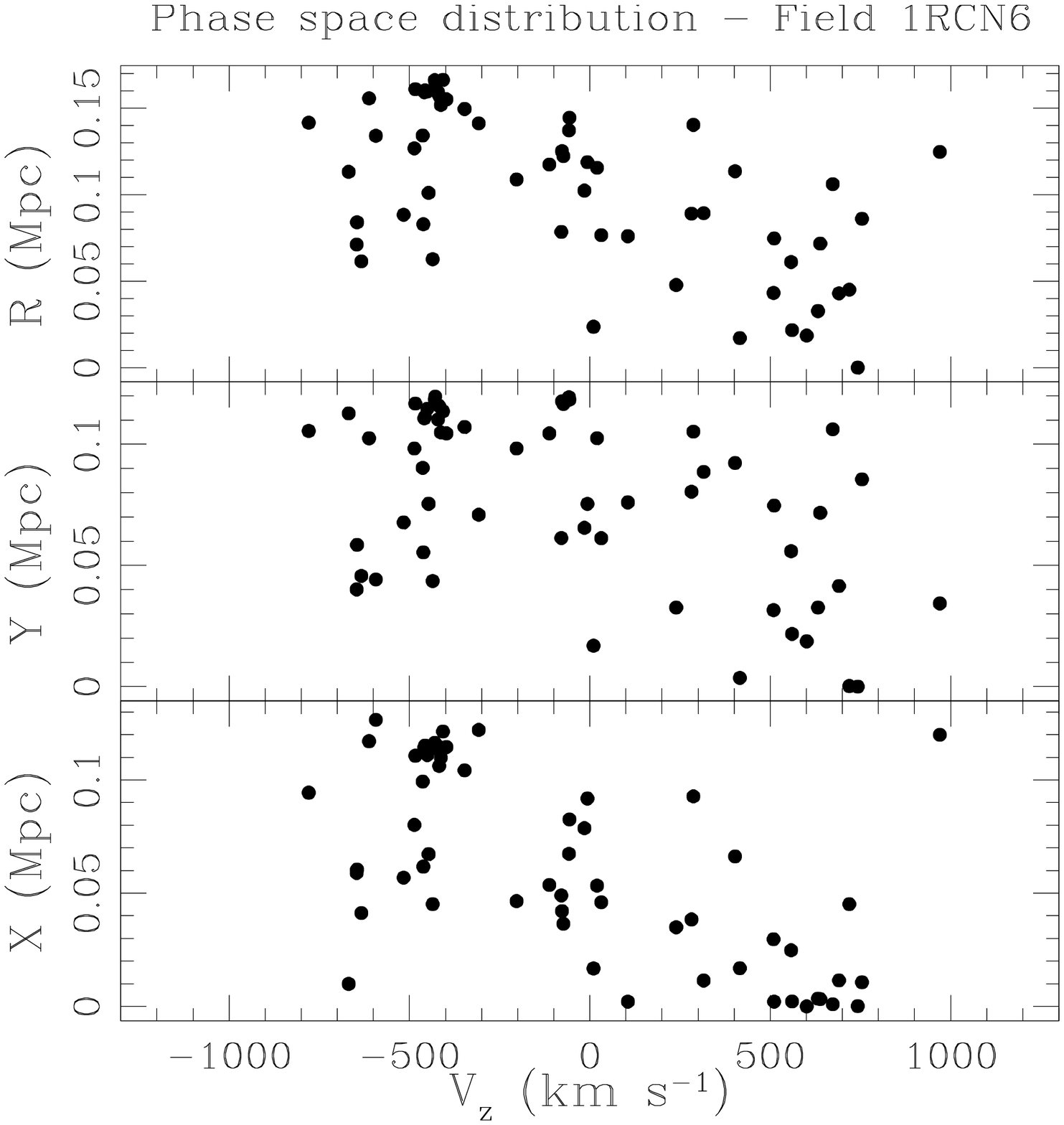,width=4.05cm,height=4.5cm}
\caption{Velocity distributions and projected phase space diagrams
for the ICSP in four RCN1--like fields.  From top to botton: velocity
distributions in transverse velocities on the sky, $V_x$ and $V_y$,
and in line-of-sight velocity $V_z$; projected phase-space diagrams
(R,$V_z$), ($x$,$V_z$), and ($y$,$V_z$), where R$=\sqrt{x^2 + y^2}$.}
\label{rcve} 
\end{figure}

{\em RCN1--like fields.} Figure~\ref{rcve} shows the velocity 
distribution of the ICSP particles in some of these fields. The mean velocity, 
standard deviation, and kurtosis for these distributions along the 
three Cartesian axes are listed in Table~\ref{tabVD}.  The inspection 
of Figure~\ref{rcve} shows clear deviations from a Gaussian 
distribution, as is also evident from the kurtosis analysis: the 
negative values of the kurtosis rule out a Gaussian distribution at 
the 95\% c.l in at least one of the Cartesian coordinates, for the 
majority of the selected fields. 

The non-Gaussian nature of the velocity distribution in
Figure~\ref{rcve} signifies that the ICSP has only partially phase
mixed. This is also seen from the lower panels of Figure~\ref{rcve}
which show projections of the particle in the RCN1 fields onto two
dimensional phase spaces. In these diagrams there are filaments,
clusters of particles, and large empty regions. The dense clump in
Field 1RCN6 comes from a local overdensity in the particle
distribution which might be a separate small subsystem. The filaments
probably originate from streams of particles which became unbound
simultaneously and still have velocities well correlated with their
positions. The voids are regions into which no such particle streams
have yet phase mixed. It is possible that these diagrams even
underestimate the amount of phase space substructures, because
individual particle orbits are often not integrated accurately over
long times in N-body codes.
 
{\em $F500$ and $F600$ fields.} Velocity distribution parameters are 
listed in Table~\ref{tabVD}. As for the RCN1--like fields, negative 
values of the kurtosis suggest a general deviation from a Gaussian 
distribution toward a flatter behavior even if it is not as strong as 
in case of the RCN1--like fields.  In some cases, in the $F5004$, 
$F5009$, $F6006$ fields, the radial velocity distributions follow a 
Gaussian. 
 
The unrelaxed, non-Gaussian nature of these velocity distributions 
shows that the ICSP has not had time to phase-mix on the 
scales probed. This is a strong prediction of this hierarchical 
cluster formation model, to be tested by future spectroscopic 
surveys.

\subsubsection{2D spatial distribution\label{2pcfRCN1}}  
 
{\em RCN1--like fields.} Using the LS estimator, we label a field as 
clustered when its $\omega(\theta)$ differs significantly from zero given the 
errors. Results are listed in Table \ref{tabclus}. 
The 2D clustering properties are anti--correlated with the degree of 
relaxation in the fields as derived by the velocity distributions: 
almost all fields that have a flat velocity distribution, in at least 
one direction, incompatible with a Gaussian distribution, are also 
clustered. 

\begin{figure} 
\epsfig{figure=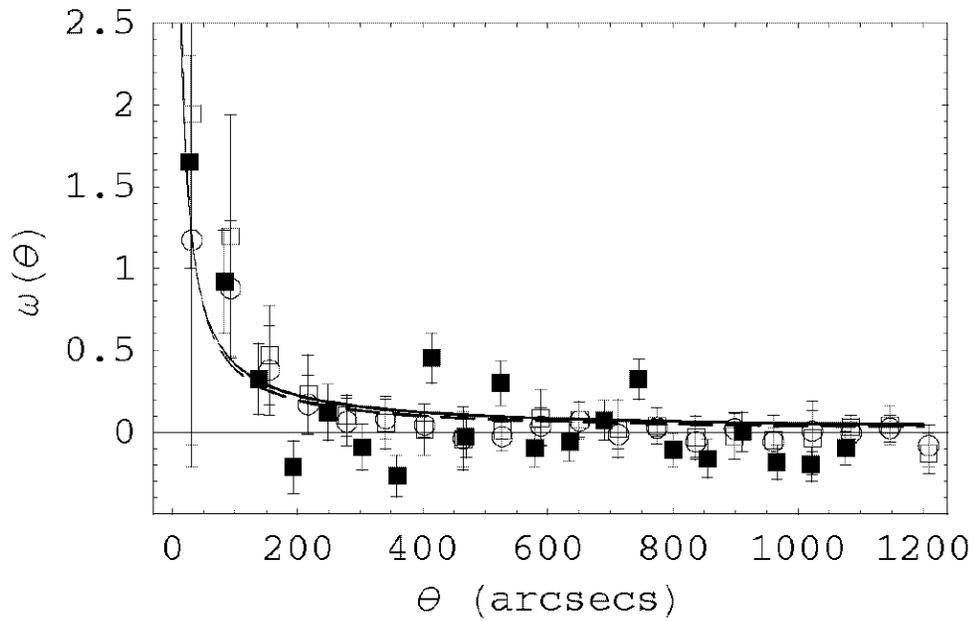,width=13cm,height=13cm} 
 
\caption{2PCFs for the RCN1 fields. Open squares are the mean $\langle 2PCF \rangle$ from all the RCN1--like fields, open circles show the mean $\langle 2PCF \rangle$ with a (25\%) contamination from a uniform population, filled squares are the 2PCF from the ICPNe data. Solid curve is the fit to open circles, dashed curve is the fit to the ICPNe sample: see discussion in the text.} 
\label{CFrcmean} 
\end{figure}

The mean 2PCF over all fields is shown in Figure~\ref{CFrcmean}, where
we compared it with that obtained from the RCN1 observed data
\citep[][]{papII}. The 2PCF from the simulation shows a stronger
clustering signal, i.e., it is systematically larger than the 2PCF from
the observed RCN1 at small $\theta$: in particular the innermost bins
have 1.5 times larger values in the simulated fields. Such an offset
can be caused \citep[see also][]{giav98} by the inferred 25\%
contamination by high-z emitters expected in the ICPNe catalogue
\citep[see][]{papI}.  In order to check this we have added a 25\%
uniform population to each of the simulated fields and computed again
the mean 2PCF for the RCN1-like fields. The results are also given in
Figure~\ref{CFrcmean}. The mean 2PCF from the simulation including the
contamination, and the ICPN 2PCF are now compatible within the
errors. The 2PCF fit to the power--law as in eq. \ref{giav98} gives
$A_w=25\pm16$ and $\beta=0.89\pm0.15$ for the average 2PCF from the
simulation fields and $A_w=35\pm20$, $\beta=0.98\pm0.31$ from the
photometrically complete sample of 55 ICPNe candidates \citep[i.e. the
sample selected within about 1 mag from the PNLF cutoff, see][for
further details]{papII}: these values are fully consistent within the
errors.  We conclude that the observed RCN1 field has a clustering
signal consistent with that expected from the N-body simulation.
 
{\em  $F500$ and $F600$ fields.\label{CFout}} 
No clear evidence for a clustering signal comes from the averaged 2PCF for both sets of fields (more marked for the $F600$ fields), except for a few $F500$ fields which are clustered. 
 
\subsection{CORE--like fields \label{corfiel}} 
 
In our simulated dataset, the BCG halo extends out to about 150 kpc 
from the galaxy center and it appears relaxed, with no evident spatial 
structures.  At a distance of 15 Mpc, 150 kpc corresponds to about 
34$'$, i.e. the distance of the observed Core field from M87. 
Therefore fields selected at these radii in the simulated data cube 
will be embedded in the BCG extended halo.  As discussed in Section~\ref{ovdens}, 
at high--z our adopted ICSP selection criteria also find DM particles. 
Thus in the extended BCG halo we expect a mix of stellar and DM tracers in our ICSP particle catalogue. 
 
{\em M/L ratios.} Because of the substantial DM component in the model 
ICSP in the CORE--like fields, the computed $\Gamma_{\mathrm{B}}$ 
values (Table \ref{tabco}) are much larger than for a pure ICSP 
component. 
 
{\em The velocity distributions.} The kurtosis values computed for the 
velocity distributions are shown in Table~\ref{tabVD}, together with 
the other parameters. Different from the RCN1--like fields, here 
negative kurtosis values are rare, and close to zero.  The large 
mass-to-light ratios and the quasi-Gaussian velocity distributions 
indicate that the selected ICSP particles are more nearly relaxed (i.e., 
phase-mixed). Furthermore, some large positive kurtosis values 
($>$1.2) suggest the presence of tails in the velocity distributions. 
 
Is this a signature of an unrelaxed population as those observed in 
the fields further away from the BCG halo?  In the BCG halo fields, 
given the setup of the simulation, we cannot say whether there is a 
{\it relaxed} ICSP component in addition to the DM particles, because we 
have no independent way of selecting it.  If the ICSP is related to 
some {\it unrelaxed} population, however, then we can hope to select it via the 
identification of substructures in the velocity distribution.  Given 
the large number of mass particles in the BCG halo fields, we can 
subtract the particles associated with a relaxed component and study 
the properties of the remaining particles.   
 
This is done as follows.  We make a Gaussian fit to the velocity 
distribution in each BCG halo field, and generate a ``relaxed'' 
population of particles whose radial velocities are randomly extracted 
from the fitted Gaussian distribution.  This relaxed population is 
then subtracted off the simulated particle distribution in the 
CORE--like fields.  
The particles remaining after this 
procedure, if any, are taken as residual ICSP population.   
In each field, we have produced 30 random realizations 
for the relaxed population from the same Gaussian radial velocity 
distribution and look for the presence of an unrelaxed 
component in all the 30 residual fields.  
Hereafter we refer to this procedure as a 
statistical velocity subtraction (SVS). 
 
\begin{figure} 
\epsfig{figure=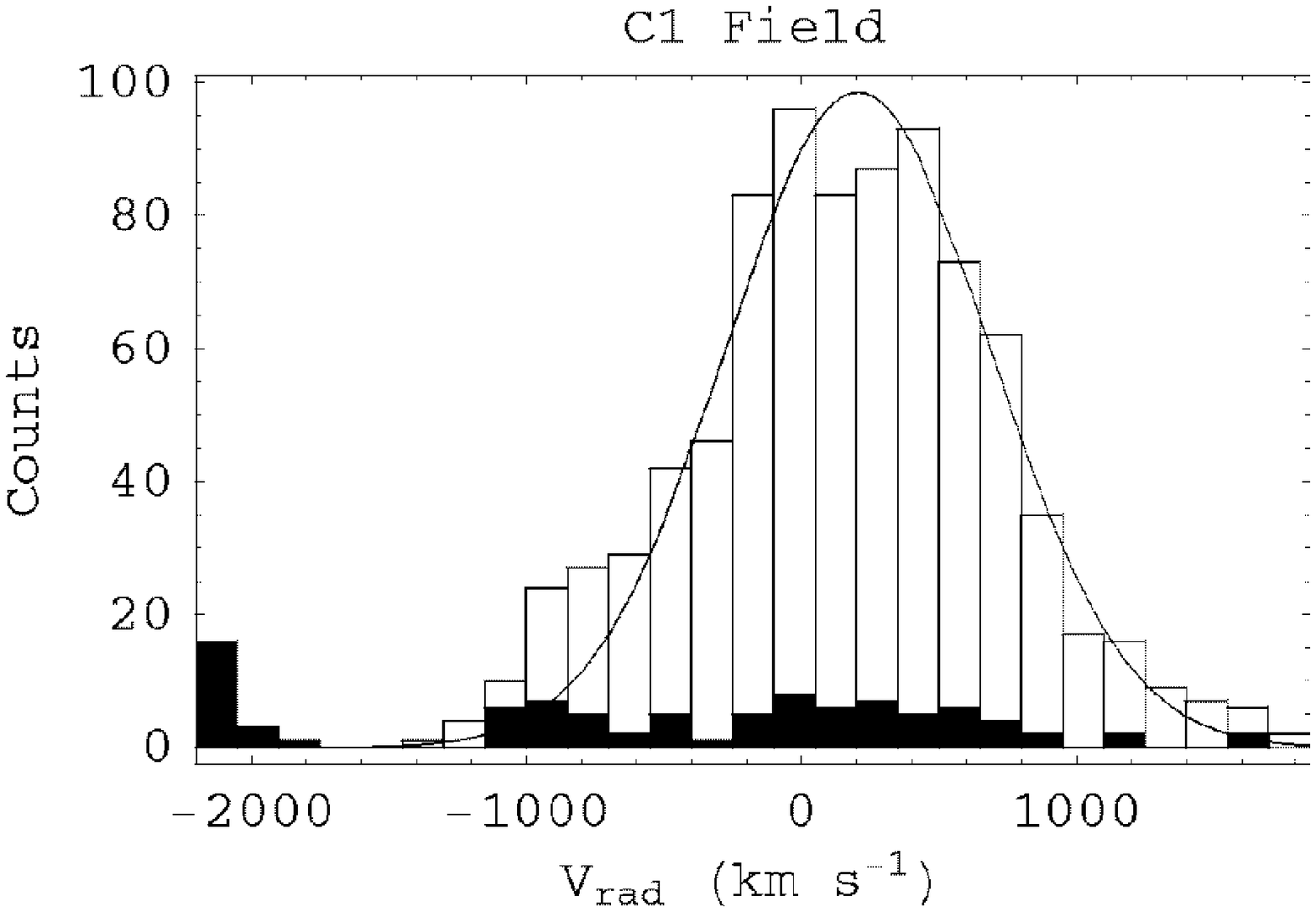,width=6.5cm,height=6.5cm} 
\hspace{-1cm} 
\epsfig{figure=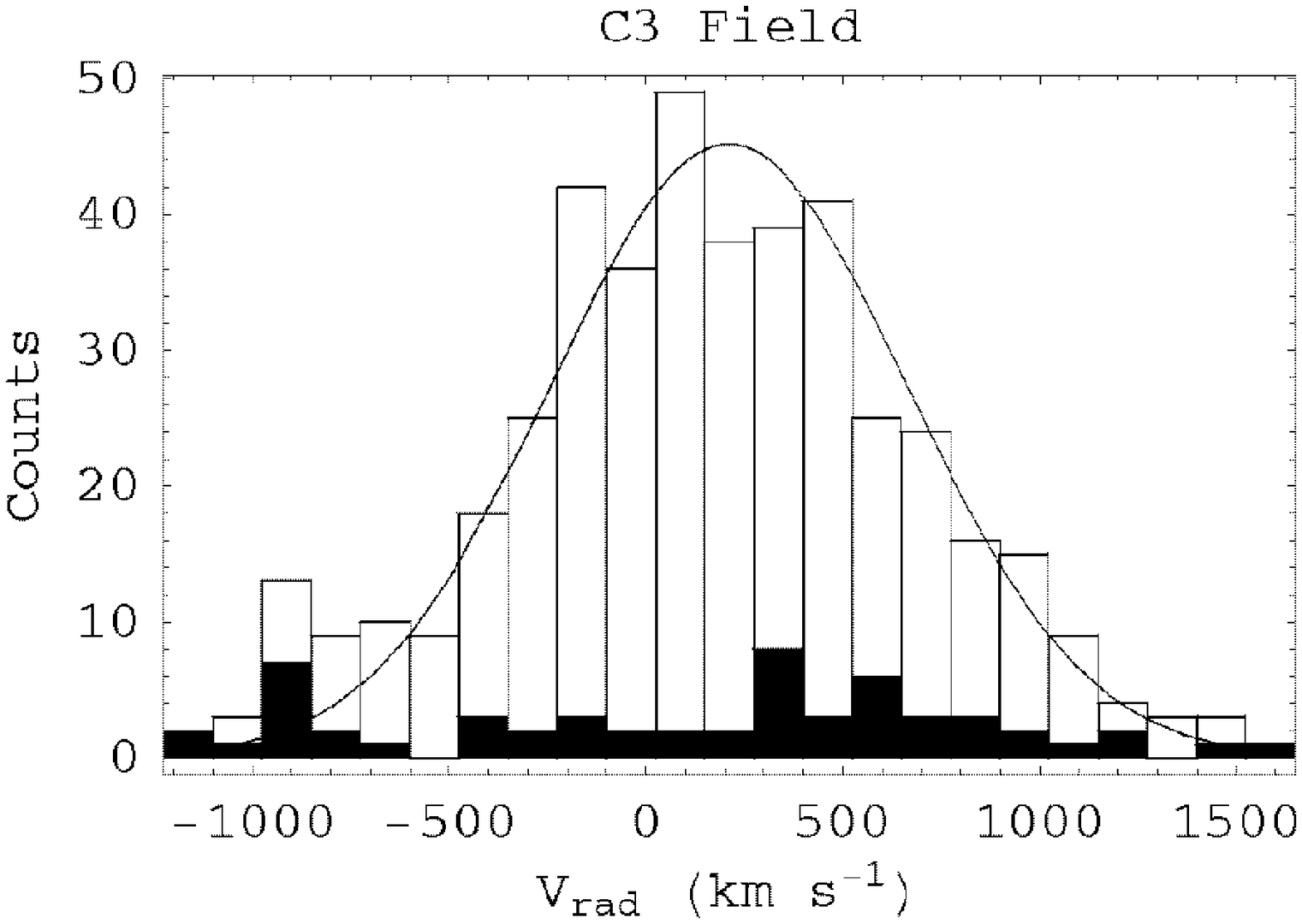,width=6.5cm,height=6.5cm} 
\epsfig{figure=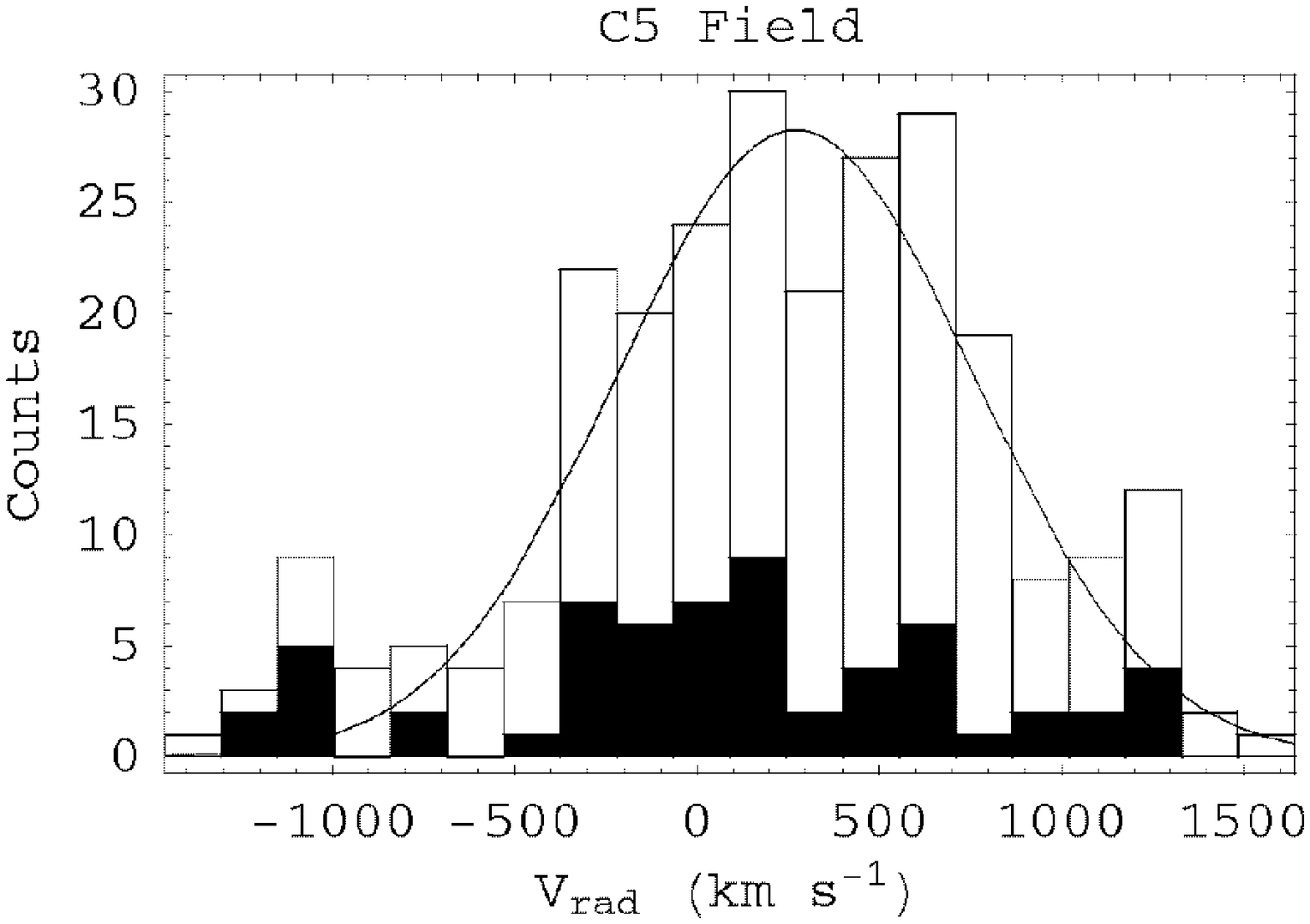,width=6.5cm,height=6.5cm} 
\hspace{+3cm} 
\epsfig{figure=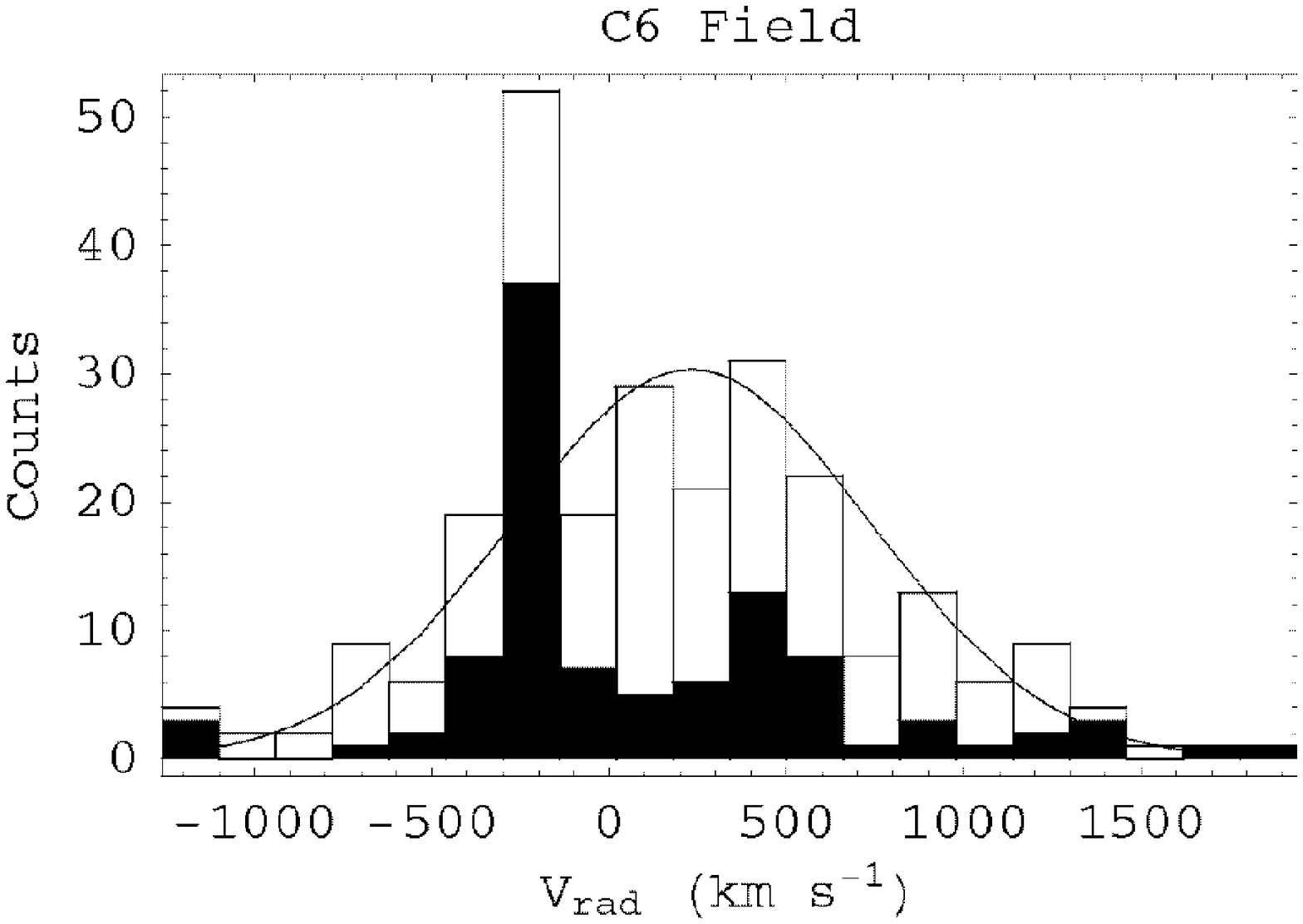,width=6.5cm,height=6.5cm} 
 
\caption{Results from the statistical velocity subtraction (SVS) for the CORE--like fields: velocity distributions of residual fields (black) in singular SVS experiments are overplotted on the velocity distributions for all the selected particles in different CORE--like fields.} 
\label{pSVS} 
\end{figure}

Are the residual particle fields in the simulated CORE--like region
compatible with the ICSP properties previously studied in the
RCN1--like fields?  Figure~\ref{pSVS} shows a sample of these residual
velocity fields; when an unrelaxed component is present the velocity
distributions are similar to those from RCN1 fields.  We can also
verify whether the M/L ratio of this unrelaxed population is
consistent with a M/L for a stellar population, and the results are
shown in Table \ref{tabSVS}.  The last row in this table lists the
averaged quantities over all Core IC fields for the residual
particles, after SVS\footnote{The field C7 has been discarded in this
analysis because it shows a double peaked Gaussian distribution due to
a couple of sub-halos which enter in the selected area: here the SVS
can give misleading results.}.  As a result, indeed the unrelaxed
component is compatible with an ICSP population and the stellar baryon
fraction, $\simeq 0.009$ in these fields is consistent with values
expected for galaxy clusters.
 
\subsubsection{Angular 2PCF \label{CF2pcf}} 
We compute the 2PCFs ($\omega(\theta)_i$, $i$=1...30) using the 
residual particle fields, then produce the mean $\langle 
\omega(\theta) \rangle$ and the relative error, given as the standard 
deviation of the $\omega(\theta)_i$ distribution. When the mean $\langle 
\omega(\theta) \rangle$ is compared with that computed for the 
all the particles in the fields the former has a 
clustering signal dominated by the DM particles, which show large 
clustering scales, while for the un-relaxed component the clustering 
scales are smaller, similar to those obtained for the RCN1--like 
fields. 
 
We have performed several tests of the SVS procedure in order to 
identify systematic effects introduced by this approach, in 
particular, how a clustering signal would be modified after the SVS 
subtraction.  As a test we have taken a RCN1--like field with 
significant clustering signal, and added a uniform distribution of 
particles, whose radial velocities where randomly extracted from a 
Gaussian with $\sigma = 400 $ kms$^{-1}$. We then applied the SVS, and 
compared the clustering signal in the residual particles with that in 
the original RCN1--like field. These tests indicate that the 
clustering signal is recovered correctly if the particles generating 
it are located in the tails of the velocity distribution, and that it 
is underestimated in the residual particles if the particles 
generating the signal are more uniformly spread over the velocity 
distribution. Applying the SVS procedure to the CORE-like fields will 
therefore not generate a spurious clustering signal. 
 
\begin{figure} 
\epsfig{figure=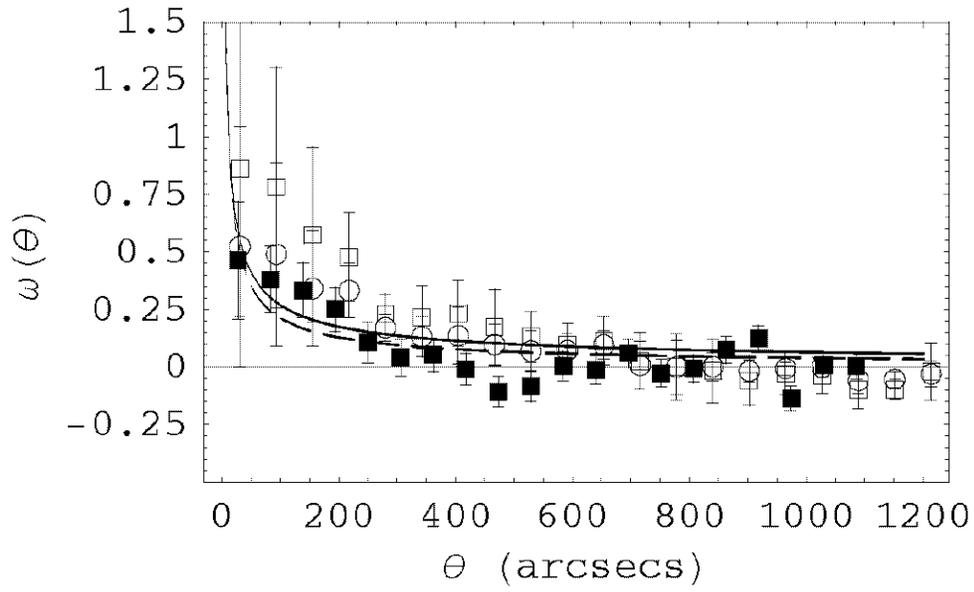,width=13cm,height=13cm} 
\caption{2PCF for the Core fields. Symbol code as in Figure \ref{CFrcmean}.} 
\label{CFcomean} 
\end{figure}

The mean 2PCF for the unrelaxed component plus a 25\% uniform sample 
(see Sect~\ref{2pcfRCN1}) in the CORE--like simulation fields is shown 
in Figure~\ref{CFcomean}, with the measured 2PCF from the observed 
ICPNe sample (77 candidates) superposed. The two are in agreement 
within the errors.  
The power law fit to the mean $\langle \omega(\theta) 
\rangle$ gives $A_w=5.2\pm2.2$ and $\beta=0.62\pm0.09$ from the 
simulated fields, and $A_w=6.1\pm4.4$, $\beta=0.73\pm0.18$ from the 
real data, considering the photometric complete sample of 77 ICPN 
candidates \citep[see][]{papII}. On the other hand, the mean 2PCF computed for all the particles in the CORE--like fields is not consistent with the 
measured 2PCF from the observed ICPN sample. 
 
The results for the CORE--like fields are indicative of the properties of an 
un-relaxed stellar-like population in these regions. The inferred M/L 
ratios and the baryonic stellar fraction are compatible with an ICSP 
population at a distance of 0.2 Mpc from the cluster center. 
Furthermore the clustering properties of this un-relaxed ICSP 
population in the CORE-like regions are in agreement with those for 
the observed ICPN data in Virgo, at the same cluster radii. 
 
\section{Global properties of the ICSP} 
 
\subsection{ICSP surface brightness and mass radial profiles at the 
 cluster scales} 
\begin{figure} 
\centering 
\vspace{-0.7cm} 
\epsfig{figure=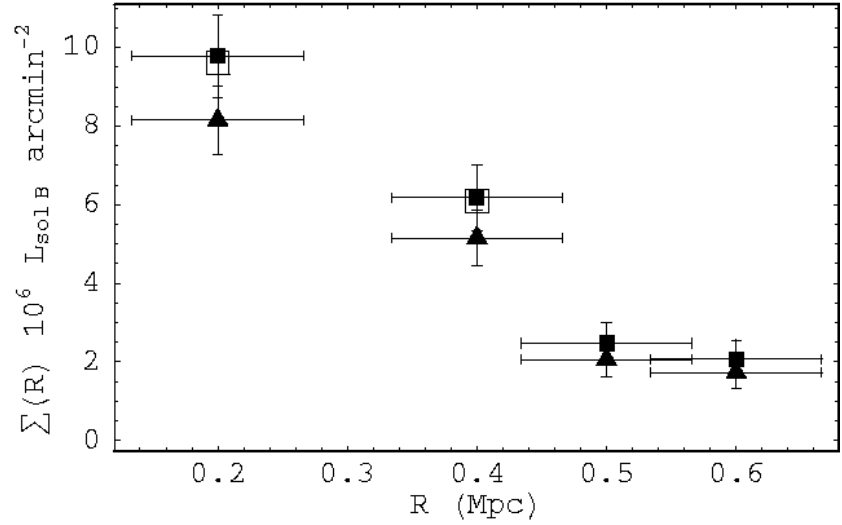,width=11cm,height=11cm} 
\epsfig{figure=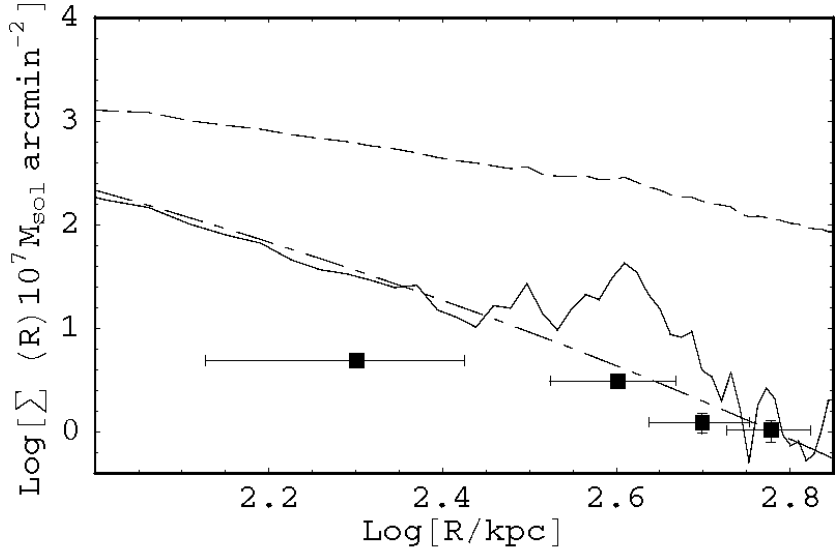,width=11cm,height=11cm} 
\vspace{-1.2cm} 
\caption{Radial surface density profiles for the ICSP. Surface brightness (SB) is on the upper panel: filled squares (triangles) are the mean SB values from the selected fields assuming $\Gamma_{\mathrm{B}}=5$ ($\Gamma_{\mathrm{B}}=6$), open squares are the estimates from the ICPNe data taking into account a 25\% contaminants in the observed ICPNe data. Surface mass density (SM) is on the lower panel: filled squares are the mean SM values from the selected fields, continuous line show the total light (i.e. mass particles selected as stellar tracers) distribution, and dashed line is the total mass (luminous+dark).} 
\label{islSB} 
\end{figure}

Having selected an ICSP in our data cube of 
simulated particles, we can now predict its radial surface density 
profile and the results are shown in Figure~\ref{islSB}. Here we 
compare the radial surface density profile from the ICSP component 
identified in the simulations with the points derived from the ICPN 
data.  In the same Figure, we plot the surface mass density for all 
the dark particles in the simulation and for all the particles 
selected as stellar tracers (cf.\ Section~\ref{selpro}).  This 
comparison allows us to estimate the fraction of the mass associated 
with the ICSP in the simulation (i.e., the unrelaxed particles in the 
BCG halo fields and all the selected particles in the Intracluster 
fields), relative to the total stellar component. The distribution of 
the latter is averaged in annuli around the cluster center before comparing it with the ICSP surface mass density in the respective fields. We find that the ICSP component accounts for 
$\sim$33\%, 29\% and 50\% of the total light in the RCN1--like, $F500$ 
and $F600$ fields respectively.  For the CORE--like fields this 
estimate provides only a lower limit, because we cannot account for 
any relaxed stellar component, which may be present at these radii as 
well.

\subsection{Clustering properties of the ICSP\label{spclus}} 
\begin{figure} 
\centering 
\vspace{-0.7cm} 
\epsfig{figure=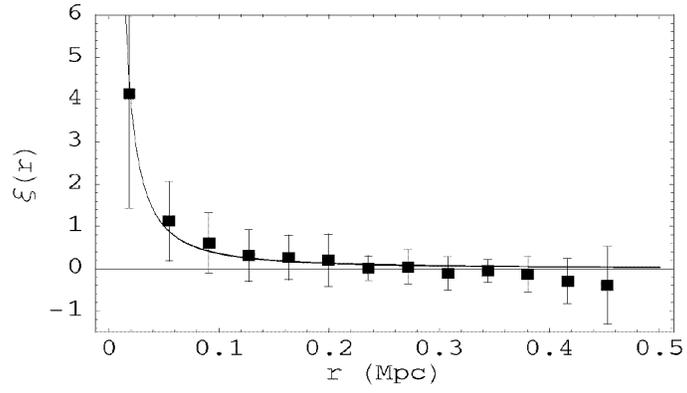,width=9cm,height=8cm} 
\vspace{-1cm} 
\caption{Average spatial 2-point correlation function from the RCN1-like fields from the simulated cluster. Error bars indicate 1 $\sigma$ confidence and solid line is the best-fit to the data. Discussion is in Section \ref{spclus}.} 
\label{spatial} 
\end{figure}

\begin{figure} 
\vspace{-1.5cm} 
\epsfig{figure=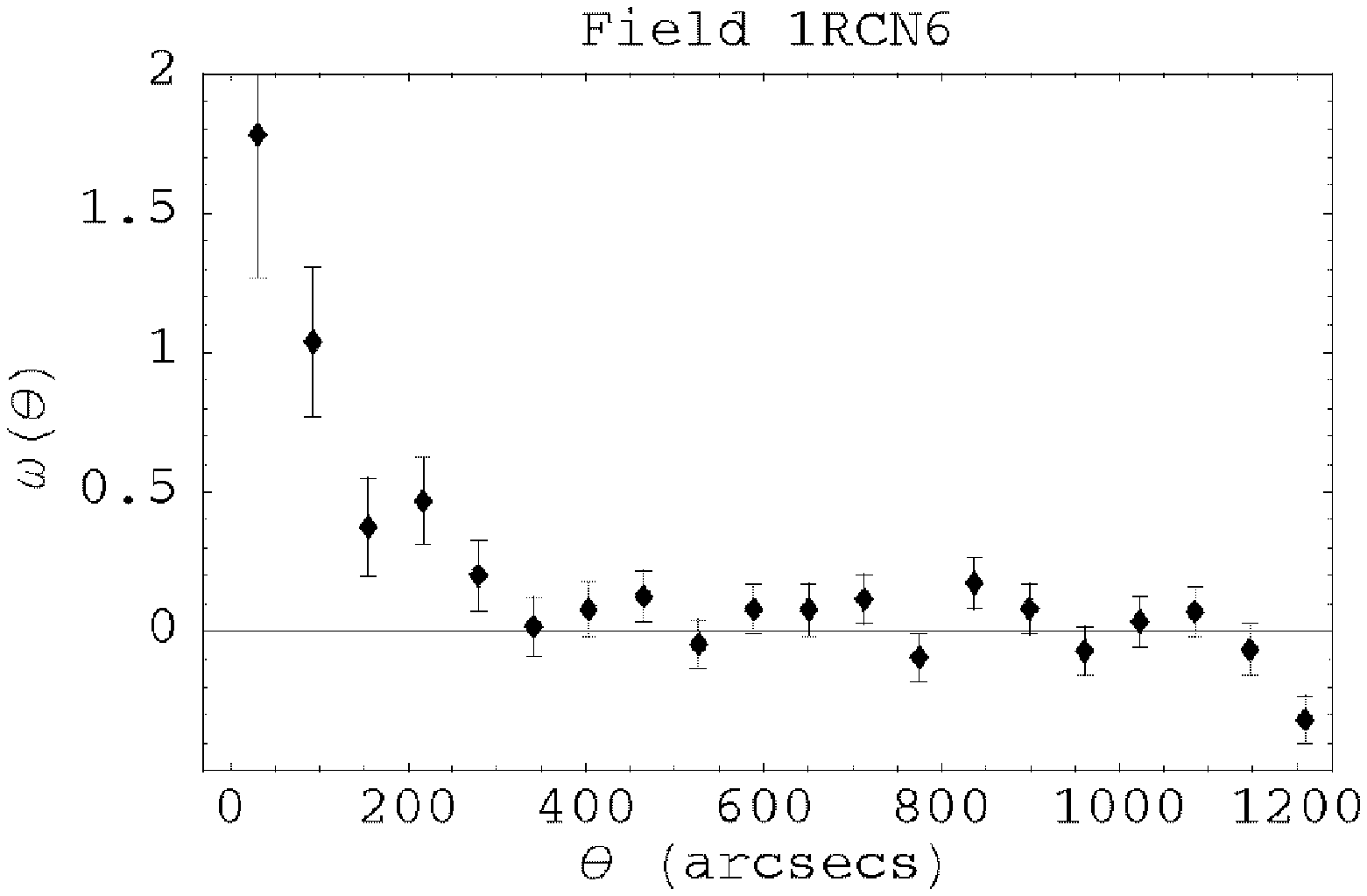,width=8cm,height=8cm} 
\epsfig{figure=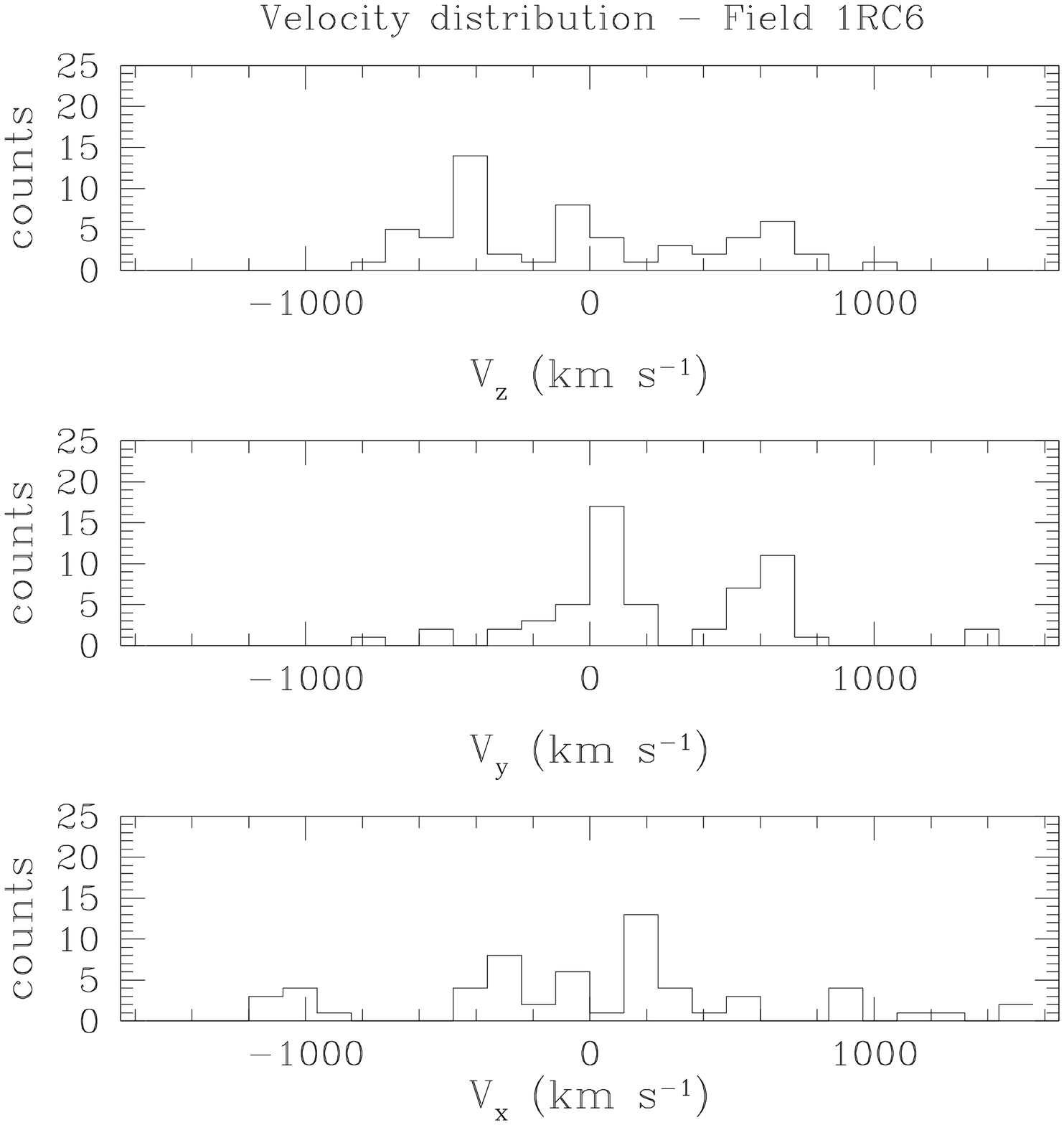,width=7cm,height=7cm} 
\epsfig{figure=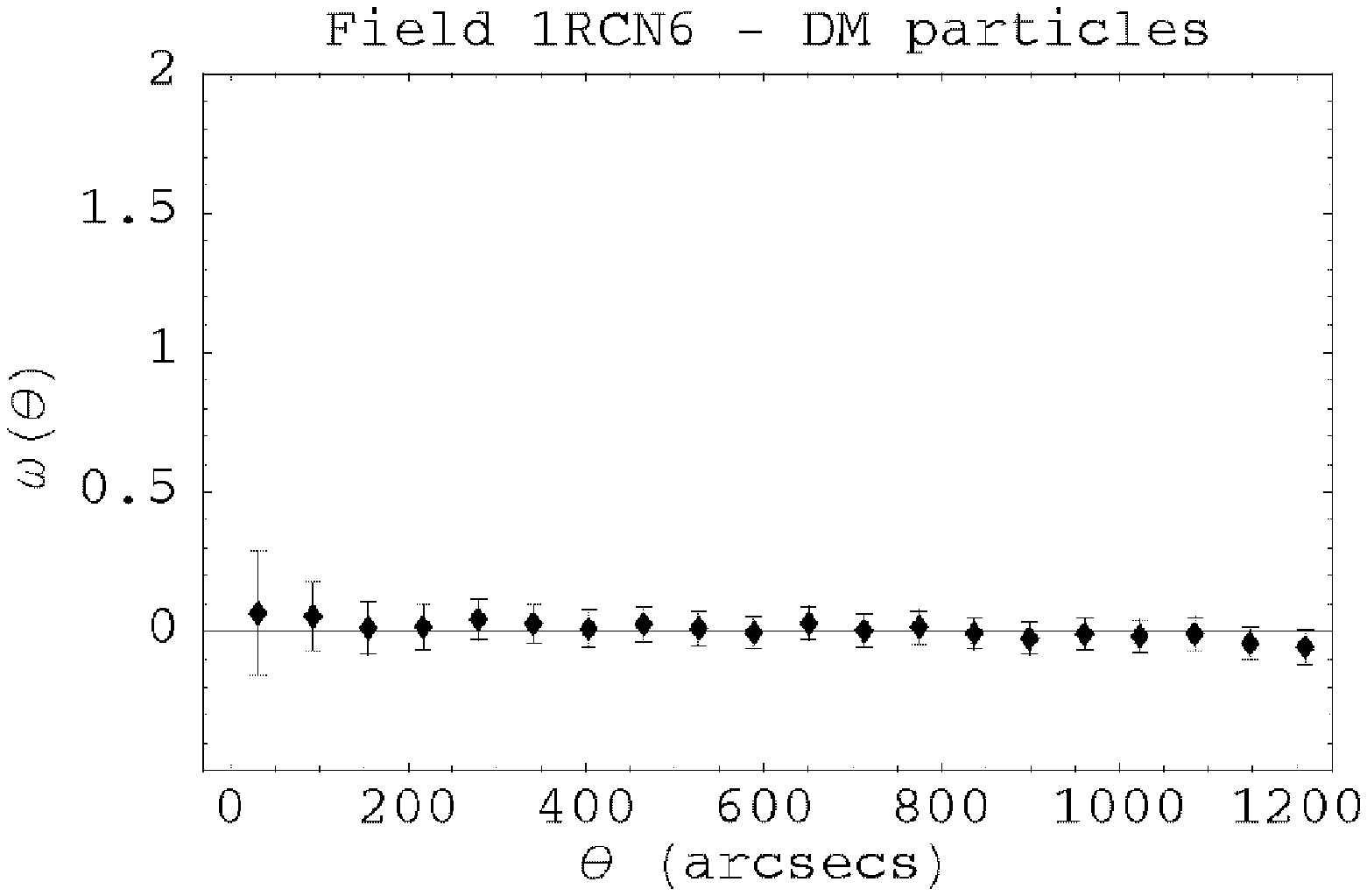,width=8cm,height=8cm} 
\hspace{+1.5cm} 
\epsfig{figure=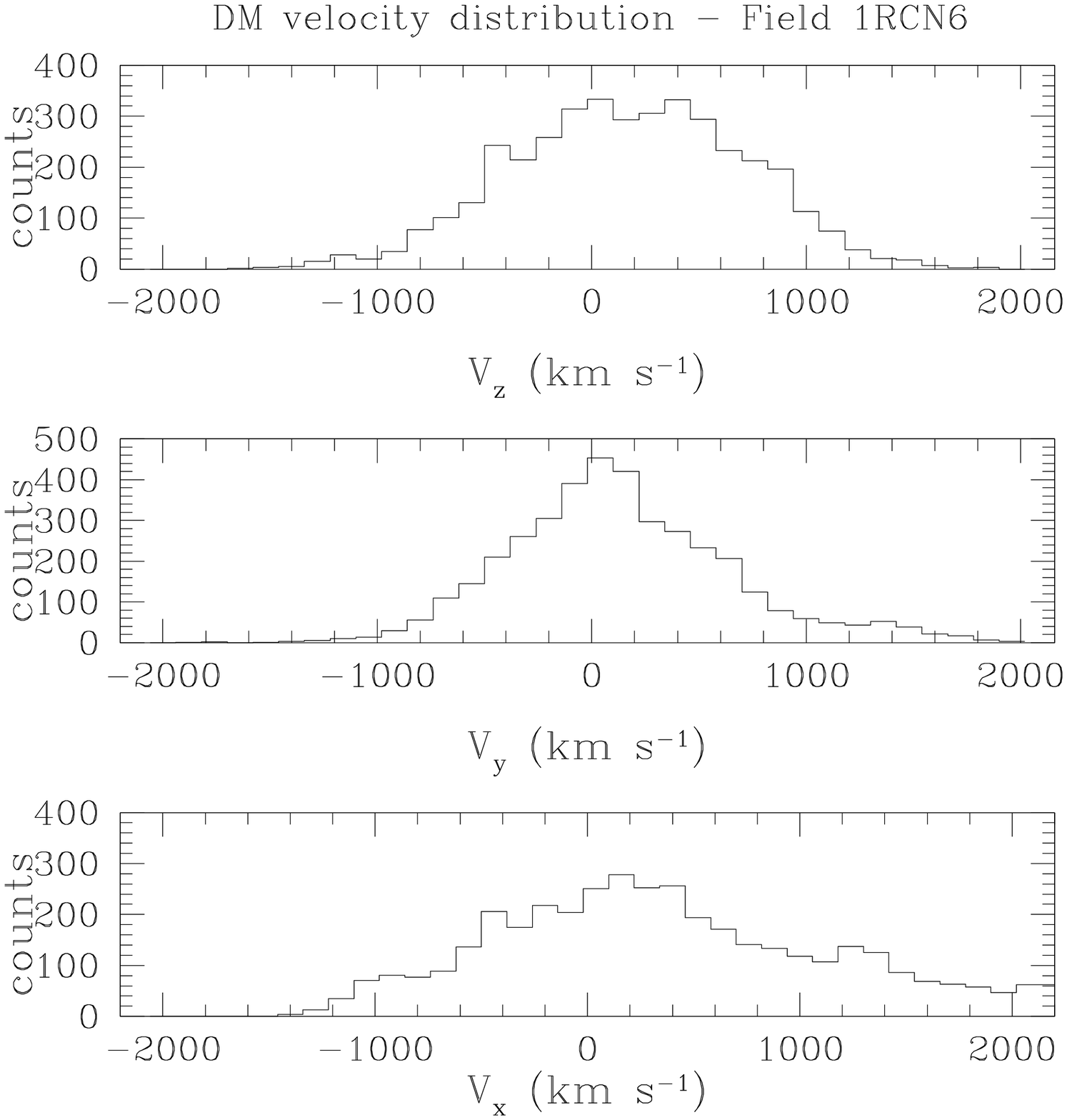,width=7cm,height=7cm} 
\caption{Observable properties of the selected stellar tracers (upper plots) vs. properties of the DM particles (lower plots) in the 1RCN6 field. Left panels: angular 2PCF; right panels: velocity distributions along the three Cartesian axes. The DM particles are relaxed and unclustered in the surveyed area, while the ICSP is clustered and unrelaxed.} 
\label{masto} 
\end{figure}

In Figures~\ref{spatial} we show the average {\sl spatial} 2PCF
$\xi(r)$ for the RCN1--like fields: it indicates the presence of
sub-structures in the ICSP. Similar features are found also in $F500$
and $F600$ fields. This evidence is stronger when we compare the
clustering properties of the selected particles with those of the DM
particles in the same field; in Figure~\ref{masto} we show that there
is no structure in the DM distribution as $ \omega(\theta)$ is zero on
all scales. A major result of this work is that the substructures in
the ICSP are also evident in the radial velocity distribution of the
ICSP particles in these IC fields, while the DM particles follow a
Gaussian distribution as shown in Figure~\ref{masto}. From a
Kolmogorov-Smirnov test, the velocity distributions for the
ICPS and DM have less than 0.001 probability to come from the
same distribution.
 
\subsection{The degree of relaxation and density environment of the 
selected particles\label{classes}}  
 
An intriguing result from our analysis is  
that those particles identified as ICSP tracers are mostly 
unrelaxed in the velocity distribution.  
The most likely interpretation is that phase-mixing did not act for  
long enough to erase streamers or tails produced during galaxy 
interactions within the cluster. 
If a stream is produced during 
encounters, this will live for a period of order several dynamical 
times, $t_{\mathrm{dyn}}$. This is a function of the mass density at 
the location where the ICSP is evolving and is given by 
\begin{equation} 
t_{\mathrm{dyn}}=\sqrt{\frac{3\pi}{16G\rho}} 
\end{equation} 
from \citet{bet87}, where $\rho$ is the mean density enclosed within 
the field distance from the center. 
Taking a mean of the selected fields at different distances we obtain 
$t_{\mathrm{dyn}}=(0.3,0.9,1.2,1.7)\times10^9$ yr for fields at 
$R=(0.2,0.4,0.5,0.6)$ Mpc respectively. Phase-mixing timescales 
are thus of the order of several gigayears. 
 
In this picture, both the degree of clustering and the velocity 
distribution of the ICSP particles give information on the history and 
evolution of the ICSP population.  Because time scales for 
phase--mixing are large in cluster environments, it is difficult, 
however, to constrain the formation epoch from such streamers or 
tails.  Perhaps more reliable constraints on the formation epoch for 
these structures may be obtained from the clustering properties which 
can be parametrised.  Using the spatial 2PCF it is possible to derive 
the correlation scales for structures in the ICSP. The parameters for 
the best fit of Equation~\ref{pippo} to the $\xi (r)$ from RCN1--like, 
F$500$ and F$600$ fields as in Figure~\ref{spatial}, are summarised in 
Table \ref{sp2P}.  Here $\gamma$ changes to lower values from the 
RCN1--like fields to the $F500$ fields, while the clustering radius 
$r_0$ becomes larger from RCN1--like to F500 fields.  
In the more distant $F600$ fields the clustering is quite weak and only present at smaller scales, as expected for 
populations of old streams for which the phase--mixing has erased the 
larger-scale structures. This suggests different dynamical evolution at different radii. 
 
In the inner regions (R$\le 0.4$ Mpc), the effects of these processes 
are evident. Here we are observing young ICSP clumps because the 
encounter probability is higher than at larger radii and structures 
are formed by harassment with higher frequency, and they are still 
unrelaxed and clustered. For structures originating more or less at 
the same epochs, the shorter dynamical time\ 
($t_{\mathrm{dyn}}^{\mathrm{0.2}}/t_{\mathrm{dyn}}^{\mathrm{0.4}}=0.37$) 
in the BCG regions decrease the clustering signal with respect to 
structures forming at R=0.4 Mpc. 
 
These estimates of $r_0$ are of the same order of magnitude as for the 
structures expected from harassment and tidal interaction in clusters 
\citep{gav01,moo96,moo98a,moo99}.  
 
The ICSP predicted in a hierarchical scenario for galaxy cluster 
formation is generally arranged in unrelaxed structures whose 
evolution is different from that of the stellar population condensed 
in galaxies. We have shown that the study of the velocity 
distributions can be used to distinguish older from younger streams. 
Constraints on the formation epoch of such a population also come from 
the clustering properties in the 2D distribution, for which 
observational data is now becoming available. Here younger streams are 
clearly identified as those with a significant clustering signal with 
respect to older streams for which clustering was erased by the random 
motions and phase-mixing.

\section{Conclusions} 
We have used an N-body cosmological simulation in a $\Lambda$CDM 
cosmology to predict the properties of the intracluster stellar 
population (ICSP) in nearby galaxy clusters.  We have selected 
particles from the simulation that trace the stellar component, by 
applying a density threshold criterion for redshifts $z\ge0.5$.  We have 
then focussed on the ICSP at $z=0$, by considering only ``intracluster'' 
fields in the model cluster, i.e., outside the central BCG galaxy and 
other cluster galaxies. The particles analysed in these fields are 
those that were lost from the inner halos they once belonged to at 
higher redshifts. 
 
We have then studied the two-dimensional spatial distribution of this
ICSP, comparing to existing data for the Virgo intracluster planetary
nebulae (ICPNe). We have considered the two-point angular correlation
function (2PCF) as a measure for the past dynamical evolution of the
ICSP component.  A significant clustering signal is expected if this
component formed relatively recently without having time to phase-mix
until the present, depending thus on the dynamical time on the
different scales.  The strongest clustering signal is found in the
inner regions of the cluster ($R \le 0.4 {\rm Mpc}$) where the
encounter probability appears to be largest. From the spatial 2PCF, we
find a correlation length of 50 kpc at R=400--500 kpc in the
simulation.  The degree of clustering seen in the N-body ICSP is in
approximate agreement with that inferred from the observed ICPNe
dataset for the Virgo cluster.

The line-of-sight velocity distributions of the ICSP particles in the
N-body model in intracluster fields are strongly non-Gaussian,
consisting often of several distinct components, which trace the
dynamical formation of the ICSP. Two-dimensional phase space diagrams
show filaments, clusters of particles, and large empty regions, all of
which indicate a young dynamical age. This is consistent with the
model ICSP being in part made of particle streams as expected in a
harassment picture.  A strong prediction of this work is that, if the
observed intracluster stars in the Virgo cluster formed in a similar
way as in this N-body model, their velocity distributions should be
highly non-Gaussian, and their phase space diagrams strongly
unrelaxed, with subcomponents that should be easily observable in
wide-field spectroscopic surveys of ICPNe.
 
Our results support the picture that the properties of the observed 
ICSP in the Virgo cluster are consistent with the hierarchical 
formation of cosmic structures and, more specifically, a complex and 
prolonged phase of galaxy infall into clusters.  In future, we will 
have larger photometric and spectroscopic ICPNe data samples to 
measure the spatial and velocity substructure of the ICSP, and more 
refined simulations, following in detail the origin from individual
harrassment events, and possibly including an explicit description of the 
star formation processes. This will allow us to compare in detail the 
internal structure of nearby galaxy clusters with cosmological cluster 
formation models, and use the properties of the ICSP to test the 
cosmological simulations and to constrain the cluster formation epoch. 
 
\acknowledgments We wish to thank the anonimous Referee for his/her
comments which helped us to improve the relevance of our results.
Simulations were run a the CINECA supercomputing center in
Bologna. NRN has received grant by the Italian Ministry of University
and Research (MIUR) (research program year 2000
ref. prot.MM02918885). OG is supported by the Swiss National Science
Foundation grant 20-64856.01. FG acknowledges partial support from the
Brooks fellowship. JALA was partially supported by the Swiss National
Science Foundation grant 20-56888.99 and by Spanish DGC (Grant
AYA2001-3939).
 
\appendix 
 
\section{Clustering via angular two--point correlation function} 
Correlations in the distribution of a sample of objects offer, perhaps, 
the simplest and least ambiguous statistical description of their 
clustering properties.  
The angular correlation function, $w(\theta)$, is the projection of the 
spatial function on the sky and is defined in terms of the joint 
probability, $\delta P$, of finding two points separated by an angular 
distance $\theta$ relative to that expected for a random distribution 
\citet{pe80}, 
 
\begin{equation} 
\delta P=N^{2}[1+w(\theta)]\delta \Omega_{1} \delta\Omega_{2}, 
\end{equation} 
 
where $\delta\Omega_{1}$ and $\delta\Omega_{2}$ are elements of solid 
angle and $N$ is the mean number density of objects.  
For an unclustered population $w(\theta)=0$, while positive or negative 
values of $w(\theta)=0$ indicate clustering or anti-clustering respectively. 
Estimators for $w(\theta)$ are generally constructed out of ratios 
between three fundamental quantities. These are the number of pairs of 
objects (DD), the number of pairs given a cross--correlation between the 
points and a random distribution (DR), and the number of pairs for a random 
distribution (RR), all suitably normalized and accounting for the geometry 
of the fields where candidates were selected.  
 
We have used the estimator proposed by \citet{ls93}: 
\begin{equation} 
w(\theta)=\frac{DD(\theta)-2DR(\theta)+RR(\theta)}{RR(\theta)}. 
\end{equation} 
The Landy \& Szalay estimator is a robust estimator because it removes edge 
effects for a field of arbitrary geometry, and gives a variance which is 
equivalent to Poissonian for an uncorrelated distribution of points.  
The angular two-point correlation function $w(\theta)$ can be represented by a power-law  
\begin{equation} 
w(\theta)=A_{w} \theta^{-\beta} 
\label{giav98} 
\end{equation} 
\citep{giav98}.  
If the spatial correlation function can be modeled as  
\begin{equation}\label{pippo} 
\xi(r)=(r/r_0)^{-\gamma}\times f(z) 
\end{equation} 
($f(z)=1$ for $z\sim 0$),  
the angular correlation function has, in fact, the form as in eq. \ref{giav98},  
where $\beta=\gamma -1$ and $r_0$ is a characteristic clustering radius. 
 
\clearpage

\clearpage 
 
\begin{deluxetable}{ccccccc} 
\tablecolumns{7} 
\tablewidth{0pc} 
\tablecaption{RCN1--like fields. (R,$\Phi$) are the polar coordinates in the sky plane, n.p.=number of particles(=ICPNe) selected in the fields, $M_t$= total mass of the particles in the fields, $f_{\mathrm{bar}}$ is the stellar baryon fraction in the selected fields. (*) $\Gamma_{\mathrm{B}}$ is obtained using the estimated surface brightness associated with the ICPNe, i.e. 6.2$\times10^6$ $L_{\odot _B}$ arcmin$^{-2}$.} 
\tablehead{ 
\colhead{Field} & \colhead{R (Mpc)}   & \colhead{$\Phi$(degrees)}    & \colhead{n.p.} & 
\colhead{$M_t(10^{10}M\odot)$}    & \colhead{$\Gamma_{\mathrm{B}} ^{(*)}$} & \colhead{$f_{\mathrm{bar}}$}}
\startdata 
 1RC1  & 0.4 &  60 & 57  & 2.885  &  5.1 &  0.014 \\
 1RC2  & 0.4 &  95 & 41  & 2.075  &  3.7 &  0.012 \\
 1RC3  & 0.4 & 220 & 137 & 6.93   & 12.4  & N/A\\
 1RC4  & 0.4 & 255 & 55  & 2.783  &  5.0  &  0.014\\
 1RC5  & 0.4 & 315 & 68  & 3.441  &  6.1  &  0.022 \\
 1RC6  & 0.4 & 350 & 58  & 2.935  &  5.2  &  0.015\\
 1RC7  & 0.4 & 143 & 49  & 2.48   &  4.4  &  0.013 \\
 1RC8 & 0.4 & 166 & 143 & 7.237  & 12.9  &  N/A\\
\enddata
\label{tabrc}
\end{deluxetable} 


\clearpage 
 
 
\begin{deluxetable}{ccccc} 
\tablecolumns{5} 
\tablewidth{0pc} 
\tablecaption{$F500$ fields. (R,$\Phi$) are the polar coordinates in the sky plane, n.p.=number of particles(=ICPNe) selected in the fields, $M_t$= total mass of the particles in the fields.} 
\tablehead{ 
\colhead{Field} & \colhead{R (Mpc)}   & \colhead{$\Phi$(degrees)}    & \colhead{n.p.} & 
\colhead{$M_t(10^{10}M\odot)$}    }
\startdata 
 $F500$1  & 0.5 &  45 & 25  & 1.265  \\
 $F500$2  & 0.5 &  60 & 21  & 1.063  \\
 $F500$3  & 0.5 & 95  & 17  & 0.860  \\
 $F500$4  & 0.5 & 140 & 21  & 1.067  \\
 $F500$5  & 0.5 & 210 & 69  & 3.492  \\
 $F500$6  & 0.5 & 240 & 25  & 1.265  \\
 $F500$7  & 0.5 & 269 & 26  & 1.316  \\
 $F500$8 & 0.5 & 310 & 20   & 1.012  \\
 $F500$8 & 0.5 & 345 & 21   & 1.067 \\
\enddata 
\label{tabr5}
\end{deluxetable} 


\clearpage 
 
\begin{deluxetable}{ccccc} 
\tablecolumns{5} 
\tablewidth{0pc} 
\tablecaption{$F600$ fields. (R,$\Phi$) are the polar coordinates in the sky plane, n.p.=number of particles(=ICPNe) selected in the fields, $M_t$= total mass of the particles in the fields.} 
\tablehead{ 
\colhead{Field} & \colhead{R (Mpc)}   & \colhead{$\Phi$(degrees)}    & \colhead{n.p.} & 
\colhead{$M_t(10^{10}M\odot)$}   }
\startdata 
 $F600$1  & 0.6 &  55 & 12  & 0.607  \\
 $F600$2  & 0.6 &  86 & 16  & 0.810  \\
 $F600$3  & 0.6 & 170 & 17  & 0.860  \\
 $F600$4  & 0.6 & 225 & 25  & 1.265  \\
 $F600$5  & 0.6 & 245 & 16  & 0.962  \\
 $F600$6  & 0.6 & 330 & 21  & 1.063  \\
 $F600$7  & 0.6 & 360 & 22  & 1.113  \\
\enddata 
\label{tabr6}
\end{deluxetable} 


\clearpage 
 
\begin{deluxetable}{cccccccccc} 
\tabletypesize{\scriptsize}
\tablecolumns{10} 
\tablewidth{0pc} 
\tablecaption{Velocity moments along the three Cartesian axes for the selected fields: mean velocity ($\overline{v_{\mathrm{i}}}$), velocity dispersion ($\sigma_{\mathrm{i}}$) and kurtosis ($k_{\mathrm{i}}$) values (i=$x, y, z$).} 
\tablehead{ 
\colhead{Field} & \colhead{$\overline{v_x}$ (km s$^{-1}$)}   & $\sigma_x$ (km s$^{-1}$) & $k_x$ & 
$\overline{v_y}$ (km s$^{-1}$)    & $\sigma_y$ (km s$^{-1}$) & $k_y$ &$\overline{v_z}$ (km s$^{-1}$) & $\sigma_z$ (km s$^{-1}$) & $k_z$}
\startdata 
 1RC1  & 80   & 401  &  -0.95  & -181 &  549 & -0.82  & -366 &  524 & 0.91  \\
 1RC2  & 279  &  259 &   -0.58 &  117 &  388 &  0.11  & 436  & 553 & -0.31  \\
 1RC3  & 96   & 447  &  -0.54  & -18  & 471  & -0.95  & 240  & 436  &  -0.46 \\
 1RC4  & 41   & 313  &   0.70  & 202  &  405 & -1.05   & 199  & 399  & 0.7  \\
 1RC5  & -274 & 477  &  3.40   & 140  &  397 & 0.84   & -16  & 333  & 0.52   \\
 1RC6  & 51   & 695  &  1.46   & 242  &  414 & 0.65   & -49  & 479  & -1.17  \\
 1RC7  & -23  & 327  &   -0.80 & 98   & 428  &  -0.22 & 5    &  384 &   -0.50 \\
 1RC8 & 50   & 449  &   -0.64 & -6   & 258  &  -0.19 & 150  &  408 &   0.5   \\
 $F500$1  & 97	& 401  & -1.02 & -93  & 301 &  2.03  & 91   & 322 &1.00  \\
 $F500$2  & -119 &  268 & -1.36 & -387 & 333 &  -0.32 & -35  & 182 & 1.45 \\
 $F500$3  & 200  & 194  & 0.12  &   0  & 476 & -0.77  & 911  & 501 &-1.53 \\
 $F500$4  & 30	& 384  & 2.59  &  125 & 255 & -0.08  & -232 & 266 &0.67  \\
 $F500$5  & 56	& 509  & -0.06 & -26  & 402 & -0.11  & 231  & 373 &-0.22 \\
 $F500$6  & 178  & 254  & -0.92 & -28  & 674 & -1.06  & 244  & 419 &-1.43 \\
 $F500$7  & 27	& 280  & -1.67 & -78  & 444 & -1.24  & -304 & 745 &-1.5  \\
 $F500$8  & -215 &  389 & -0.44 &  125 & 204 & 1.08   & -1   & 263 &2.54  \\
 $F500$9  & 100  & 441  & 1.46  & -68  & 436 & 0.37   & 187  & 321 &1.47  \\
 $F600$1  & -16 & 154 & -1.03 & -18 & 203 & -0.20  & -337 & 336 & -0.44 \\
 $F600$2  & 337 & 165 & -0.36 & -300 & 605 & -0.78 & 316 & 367 & 2.62  \\
 $F600$3  & 395 & 681 & -0.45 & 122 & 358 & 4.96  & 366 & 561 & -1.37  \\
 $F600$4  & -32 & 247 & -0.07 & -109 & 321 & -0.95 & 51 & 360 & -1.53  \\
 $F600$5  & 163 & 219 & 5.39 & 141 & 409 & -0.53 & 372 & 186 & 0.89   \\
 $F600$6  & -95 & 428 & 1.10 & 38 & 235 & 0.30 & 116 & 236 & 2.77  \\
 $F600$7  & 894 & 677 & -0.54 & 53  & 209  & 0.35 & 44 & 277 & -0.12 \\
 C1  & 122 &  474 &  2.5  &  151 &   459 & 0.09   &   107 &   643 & 0.09     \\
 C2  & 127 &  551 &  1.21 &  114 &   603 &  0.28  &   122 &   540 &  0.28    \\
 C3  & 173 &  401 &  3.92 &  117 &   534 &  -0.35 &   161 &   522 &  -0.35   \\
 C4  & 42  & 424  & 1.57  & 87   & 480   & 0.39   & 185   & 568   & 0.39     \\
 C5  & 24  & 432  & 1.21  & 127  &  541  & -0.19  &  221  &  591  & -0.19    \\
 C6  & -2  & 605  & -0.33 & 230  &  419  &  0.24  &  162  &  555  &  0.24    \\
 C7  & 301 &  767 & 0.63  &  172 &   336 & 1.98   &   121 &   455 & 1.98     \\
\enddata 
\label{tabVD}
\end{deluxetable} 


\clearpage 
 
\begin{deluxetable}{cc} 
\tablecolumns{2} 
\tablewidth{0pc} 
\tablecaption{Clustering of the RCN1-like fields.} 
\tablehead{ 
\colhead{Field} & \colhead{CLASS} }
\startdata 
 1RC1   &  clustered    \\
 1RC2   &  clustered    \\
 1RC3   &  clustered    \\
 1RC4   &  clustered    \\
 1RC5   &  clustered \\
 1RC6   &  clustered    \\
 1RC7   &  not clustered \\
 1RC8   &  not clustered \\
\enddata 
\label{tabclus}
\end{deluxetable} 
%

\clearpage 
 
\begin{deluxetable}{cccccc} 
\tablecolumns{6} 
\tablewidth{0pc} 
\tablecaption{CORE--like fields. (R,$\Phi$) are the polar coordinates in the sky plane, n.p.=number of particles(=ICPNe) selected in the fields, $M_t$= total mass of the particles in the fields. (*) $\Gamma_{\mathrm{B}}$ is obtained using the estimated surface brightness associated with the ICPNe, i.e. 9.7$\times10^6$ $L_{\odot _B}$ arcmin$^{-2}$.} 
\tablehead{ 
\colhead{Field} & \colhead{R (Mpc)}   & \colhead{$\Phi$(degrees)}    & \colhead{n.p.} & 
\colhead{$M_t(10^{11}M\odot)$}    & \colhead{$\Gamma_{\mathrm{B}} ^{(*)}$}}
\startdata 
 C1  & 0.2 & 140 & 872 & 4.413   & 51.2  \\
 C2  & 0.2 &  70 & 535 & 2.707   & 31.4  \\
 C3  & 0.2 & 105 & 434 & 2.196   & 25.5  \\
 C4  & 0.2 & 240 & 734 & 3.7145  & 43.1  \\
 C5  & 0.2 & 280 & 257 & 1.3001  & 15.1  \\
 C6  & 0.2 & 310 & 259 & 1.3107  & 15.2  \\
 C7  & 0.2 &   7 & 754 & 3.8156  & 44.3  \\
\enddata 
\label{tabco}
\end{deluxetable} 

\clearpage 
 
\begin{deluxetable}{cccccc} 
\tablecolumns{6} 
\tablewidth{0pc} 
\tablecaption{Results for CORE--like fields, after the SVS. $\langle V \rangle$=mean and SD= standard deviation of the gaussian fits; a.n.p.= the average number of objected selected in the SVS (over 30 experiments). Last row are the mean values over the six fields. (*) $\Gamma_{\mathrm{B}}$ is obtained using the estimated surface brightness associated with the ICPNe, i.e. 9.7$\times10^6$ $L_{\odot _B}$ arcmin$^{-2}$.} 
\tablehead{ 
\colhead{Field} & \colhead{$\langle V \rangle$}   & \colhead{SD}    & \colhead{a.n.p.} & 
\colhead{$M_t(10^{10}M\odot)$}    & \colhead{$\Gamma_{\mathrm{B}} ^{(*)}$}}
\startdata 
 C1  & 206 & 483 & 93 & 4.710  & 5.5  \\
 C2  & 232 & 421 & 94 & 4.757  & 5.5  \\
 C3  & 211 & 452 & 71 & 3.593  & 4.2  \\
 C4  & 205 & 488 & 99 & 5.010  & 5.8  \\
 C5  & 271 & 491 & 67 & 3.391  & 4.0  \\
 C6  & 232 & 515 & 96 & 4.858  & 5.6  \\
 $\langle C \rangle$ &     &     & 87 & 4.387  & 5.0  \\
\enddata 
\label{tabSVS}
\end{deluxetable} 


\clearpage 
 
\begin{deluxetable}{ccc} 
\tablecolumns{3} 
\tablewidth{0pc} 
\tablecaption{Mean spatial 2PCF outside the cD halo: parameters of the fits with 1 $\sigma$ errors.} 
\tablehead{ 
\colhead{Field} & \colhead{$\gamma$} & \colhead{$r_0 (kpc)$} }
\startdata 
 RCN1   &  $1.49 \pm 0.16$  &  $50\pm 5$\\
 $F500$   &  $1.35 \pm 0.18$  &  $58\pm 7$ \\
 $F600$   &  $1.8 \pm 0.9$  &  $30\pm 10$ \\
\enddata 
\label{sp2P}
\end{deluxetable} 
%

\end{document}